\pdfoutput=1
\newcommand*{\ATLASLATEXPATH}{latex/}
\documentclass[UKenglish,cernpreprint,texlive=2016,txfonts,LANGEDIT=true]{\ATLASLATEXPATH atlasdoc}

\usepackage[title]{appendix}
\usepackage[backend=biber]{\ATLASLATEXPATH atlaspackage}
\usepackage{\ATLASLATEXPATH atlasbiblatex}
\usepackage{\ATLASLATEXPATH atlascontribute}
\usepackage{\ATLASLATEXPATH atlasphysics}
\usepackage{\ATLASLATEXPATH atlasheavyion}
\usepackage{\ATLASLATEXPATH atlasunit}
\usepackage{siunitx}

\graphicspath{{logos/}{figures/}}
\usepackage{HION-2017-11-PAPER-defs}

\AtlasTitle{
Measurement of long-range two-particle azimuthal correlations in 
  $Z$-boson tagged $pp$ collisions at $\sqrt{s}{=}8$ and 13~TeV
}

\AtlasAbstract{%
Results are presented from the measurement by ATLAS of long-range ($|\Delta\eta|>2$)
  dihadron angular correlations in $\sqrt{s}=8$ and 13~TeV $pp$ 
  collisions containing a $Z$ boson. 
The analysis is performed using 19.4~$\mathrm{fb}^{-1}$ of 
  $\sqrt{s}=8$~TeV data recorded during Run 1 of the LHC and 
  36.1~$\mathrm{fb}^{-1}$ of $\sqrt{s}=13$~TeV data recorded during Run 2. 
Two-particle correlation functions are measured as a function of 
  relative azimuthal angle over the relative pseudorapidity 
  range $2<|\Delta\eta|<5$ for different intervals of charged-particle 
  multiplicity and transverse momentum. 
The measurements are corrected for the presence of background charged 
 particles generated by collisions that occur during one passage of two colliding proton bunches in the LHC. 
Contributions to the two-particle correlation functions from hard 
  processes are removed using a template-fitting procedure. 
Sinusoidal modulation in the correlation functions is observed and 
  quantified by the second Fourier coefficient of the correlation 
  function, $v_{2,2}$, which in turn is used to obtain the 
  single-particle anisotropy coefficient $v_{2}$.  
The $v_{2}$ values in the $Z$-tagged events, integrated over 
  $0.5<p_{\mathrm{T}}<5$~GeV,  
  are found to be independent of multiplicity and $\sqrt{s}$, and consistent within 
  uncertainties with previous measurements in inclusive 
  $pp$ collisions. 
As a function of charged-particle $p_{\mathrm{T}}$, the $Z$-tagged and 
  inclusive $v_{2}$ values are consistent within uncertainties for 
  $p_{\mathrm{T}}<3$~GeV.
}

\AtlasRefCode{HION-2017-11}

\PreprintIdNumber{CERN-EP-2019-103}

\AtlasDate{\today}

\AtlasJournalRef{Eur. Phys. J. C 80 (2020) 64}
\AtlasDOI{10.1140/epjc/s10052-020-7606-6}

\hypersetup{pdftitle={ATLAS draft},pdfauthor={The ATLAS Collaboration}}

\newcommand{\papertype}{paper}
\newcommand{\figdir}{figs/}

\begin{document}

\maketitle
\section{Introduction\label{sec:introduction}}
Measurements of two-particle correlations~(2PC) in relative azimuthal
  angle,
  $\dphi=\phi^{{\mathrm{a}}}-\phi^{{\mathrm{b}}}$, and
  pseudo-rapidity separation\footnote{The labels $a$ and $b$ denote the two particles in the pair.}
  $\deta=\eta^{{\mathrm{a}}}-\eta^{{\mathrm{b}}}$ in
  proton--proton~(\pp) collisions show the presence of correlations in
  \dphi at large $\eta$
  separation~\cite{CMS-QCD-10-002,HION-2015-09,HION-2016-01,CMS-HIN-16-010}.\footnote{\AtlasCoordFootnote}
Recent studies by the ATLAS Collaboration demonstrate that these
  long-range correlations  are consistent with the presence of a
  cosine modulation of the single-particle azimuthal angle
  distributions~\cite{HION-2015-09,HION-2016-01}, similar to that
  seen in nucleus-nucleus (A+A)~\cite{HION-2011-01,HION-2011-05,HION-2016-06,HION-2019-03,Aamodt:2010pa,CMS-HIN-10-002,Adams:2005dq,Back:2004je,Adcox:2004mh,Arsene:2004fa}
  and proton-nucleus (\pA) collisions~\cite{HION-2016-01,CMS-HIN-12-005,Abelev:2012ola,HION-2012-13,HION-2013-04,Aaij:2015qcq,Acharya:2019vdf}.
The modulation of the single-particle azimuthal angle distributions
  is typically characterized using a set of Fourier coefficients
  \vn, also called flow harmonics, that describe the relative amplitudes of the sinusoidal
  components of the single-particle distributions:
\begin{eqnarray}
\frac{{\mathrm{d}}N}{{\mathrm{d}}\phi}\propto\left(1+2\sum_{n=1}^{\infty}v_{n}\cos\Bigl(n(\phi-\Phi_{n})\Bigr)\right),
\label{eq1}
\end{eqnarray}
where the \vn and $\Phi_{n}$ denote the magnitude and orientation
  of the $n^{\mathrm{th}}$-order single-particle anisotropies.

The \vn in \NucNuc\ collisions result from anisotropies of the initial collision geometry,
  which are subsequently transferred to the azimuthal distributions of
  the produced particles by the collective evolution of the medium.
This transfer of the spatial anisotropies in the initial collision
  geometry to anisotropies in the final particle distributions is well
  described by relativistic hydrodynamics~\cite{Sorge:1996pc,Huovinen:2001cy,Kolb:2001qz,
                                                Gale:2013da,Heinz:2013th}.
The ATLAS measurements\,\cite{HION-2016-01} show that the \pT\
  dependence of the second-order harmonic, \vtwo, in \pp\ collisions
  is similar to the dependence observed in \pA\ and \NucNuc\ collisions.
Additionally, the $\vtwo(\pt)$ in \pp\ collisions shows no dependence on
  the centre-of-mass collision energy, \sqs, from 2.76~\TeV\ to 13~\TeV,
  similar to what is observed in \pA and \NucNuc collisions~\cite{HION-2011-05,HION-2011-01,HION-2016-06}.
 The observation that the \pT and \sqs dependences of \vtwo are each strikingly similar between \pp\ collisions and \pA and \NucNuc collisions indicates
 the possibility of collective behaviour developing in \pp\
  collisions, although alternative models 
  exist that qualitatively reproduce the features observed in the 
  \pp\ 2PC~\cite{Dumitru:2010iy,Kovner:2010xk,Altinoluk:2011qy,
                 Dusling:2012iga,Levin:2011fb,Strikman:2011cx,
                 He:2015hfa,Romatschke:2015dha,Kurkela:2018ygx}.

One feature in which the \pp\ \vtwo differs from the \vtwo in 
  \NucNuc collisions is that the \pp\ \vtwo is observed to be
  independent, within uncertainties, of the event multiplicity~\cite{HION-2015-09,HION-2016-01},
  while the \NucNuc\ \vtwo exhibit considerable
  dependence on the event multiplicity~\cite{HION-2011-01,HION-2011-05,HION-2016-06,HION-2019-03}.
This dependence is understood to be due to a correlation between the
  collision geometry and collision impact parameter (\bimp)~\cite{Snellings:2011sz}.
In collisions with small \bimp the second-order eccentricity
  $\epsilon_{2}$~\cite{Qin:2010pf,HION-2012-10} quantifying the
  ellipticity of the initial collision geometry is small, resulting
  in a small \vtwo.
Interactions at $\bimp\sim R$, where $R$ is the nuclear radius,
  result in an overlap region that becomes increasingly elliptic,
  with $\epsilon_{2}$ increasing with \bimp.
This, in turn, generates larger \vtwo.
Thus, the strong correlation between the \vtwo and multiplicity is in
  fact the result of the dependence of the collision geometry on \bimp.
There are multiple theoretical studies in \pA and \NucNuc
  collisions which reproduce the \bimp dependence of the \vn quite well~\cite{Gale:2013da}.
However, there are very few such calculations for \pp collisions.
A recent study, that models the proton substructure that can induce
  event-by-event fluctuations in the number of final particles, showed that the
  eccentricities $\epsilon_{2}$ and $\epsilon_{3}$ of the initial
  entropy-density distributions in \pp collisions have no correlation
  with the final particle multiplicity~\cite{Welsh:2016siu}.

This \papertype\ reports the long-range correlations of charged particles
  measured in \pp\ interactions  that contain a \Zboson boson
  decaying to dimuons.
The presence of a \Zboson boson selects events in which a
  hard scattering with momentum transfer Q$^2\gtrapprox(80\,\mathrm{\GeV})^{2}$ occurred.
Based on the arguments in Ref.~\cite{Frankfurt:2003td}, such events on average
  may have a lower impact parameter, \bimp, than \pp\ events without
  any requirement on Q$^2$ (termed \textit{inclusive} events in this \papertype).
An assumption, driven by the measurements performed in \NucNuc collisions, is that if the
  \pp\ \vtwo is related to the eccentricity of the collision geometry,
  then events `tagged' by a \Zboson boson having a smaller \bimp might
  also have a smaller \vtwo value than that measured in inclusive events.
As in previous ATLAS analyses of long-range correlations in \pPb\ and
  \pp collisions~\cite{HION-2012-13,HION-2013-04,HION-2015-09,HION-2016-01},
  the measured charged-particle multiplicity, uncorrected for
  detector efficiency, is used to quantify the event activity.

The data used in previous ATLAS \pp\ studies investigating structures observed
  in the long-range two-particle correlations,  also known as `ridge'~\cite{HION-2015-09,HION-2016-01},
  were recorded under conditions of low instantaneous
  luminosity, for which the number of collisions per bunch crossing
  (\nint), was $\nint\lesssim{1}$.
However, the \Zboson-boson dataset used in the present analysis is
  characterized by significantly higher luminosity conditions, with a typical
  \nint of about 20.
This large luminosity poses significant complications to the correlation
  analysis, as it is not possible to fully separate reconstructed tracks
  associated with the interaction producing the \Zboson boson
  from tracks from other interactions (pile-up) in the same bunch crossing.
In order to solve the problem of pile-up tracks, a new procedure is developed that on a
  statistical basis corrects the multiplicity and removes the contribution of
  pile-up tracks from the measured 2PC.

The \papertype\ is organized as follows.
Section\,\ref{sec:atlas2} gives a brief overview of the ATLAS detector
  subsystems.
Section\,\ref{sec:Datasets_trigger_and_tracking_cuts} describes the
  dataset, triggers and the offline selection criteria used to select
  events and reconstruct charged-particle tracks used in the analysis.
Section\,\ref{sec:2PC} gives a brief overview of the two-particle
  correlation method and how it is used to obtain the \vtwo.
Section\,\ref{sec:pile-up} details the corrections applied for analysing
  data in the presence of background from pile-up.
In Section\,\ref{sec::template_fits}, the two-particle correlations are
  calculated following procedures described in Refs.\,\cite{HION-2015-09,HION-2016-01}.
The systematic uncertainties are detailed in Section\,\ref{sec:syst}
  and the results are presented and discussed in Section\,\ref{sec:results}.
Section~\ref{sec:summary} gives the summary.

\section{ATLAS detector\label{sec:atlas2}}

The ATLAS detector~\cite{PERF-2007-01} at the LHC covers nearly the entire solid angle 
  around the collision point. 
It consists of an inner tracking detector surrounded by a thin superconducting solenoid, 
  electromagnetic and hadronic calorimeters, and a muon spectrometer incorporating three 
  large superconducting toroid magnets. 
The inner-detector system (ID) 
consisting of a silicon pixel detector, a silicon microstrip tracker and a transition radiation tracker
is immersed in a \SI{2}{\tesla} axial magnetic field. 
The ID provides charged-particle tracking in the range $|\eta| < 2.5$.

The high-granularity silicon pixel detector covers the \pp interaction region and 
  typically provides three measurements per track. 
In the 13~\TeV\ data samples, the number of measurements per track is 
  increased to four because  an additional silicon layer, 
	the insertable B-layer (IBL) detector~\cite{ATLAS-TDR-2010-19,Abbott:2018ikt},  was installed prior to the 13~\TeV\ data-taking. 
The pixel detector is followed by the silicon microstrip tracker, which 
  typically provides measurements of four two-dimensional points per track. 
These silicon detectors are complemented by the transition radiation 
  tracker, which enables radially extended track reconstruction up to 
	$|\eta| = 2.0$, providing around 30 hits per track.  

The calorimeter system covers the pseudorapidity range $|\eta| < 4.9$.
Within the region $|\eta|< 3.2$, electromagnetic calorimetry is provided by barrel and 
endcap high-granularity lead/liquid-argon (LAr) electromagnetic calorimeters,
with an additional thin LAr presampler covering $|\eta| < 1.8$,
to correct for energy loss in the material upstream of the calorimeters.
Hadronic calorimetry is provided by a steel/scintillating-tile calorimeter,
segmented into three barrel structures within $|\eta| < 1.7$, and two copper/LAr hadronic endcap calorimeters.
The solid angle coverage is completed with forward copper/LAr and tungsten/LAr calorimeter modules
optimized for electromagnetic and hadronic measurements respectively.

The muon spectrometer (MS) comprises separate trigger and
high-precision tracking chambers measuring the deflection of muons in a magnetic field generated by superconducting air-core toroids.
The precision chamber system covers the region $|\eta| < 2.7$ with three layers of monitored drift tubes,
complemented by cathode strip chambers in the forward region, where the background is highest.
The muon trigger system covers the range $|\eta| < 2.4$ with resistive plate chambers in the barrel, and thin gap chambers in the endcap regions.

A multi-level trigger system is used to select 
events of interest for recording~\cite{PERF-2011-02,TRIG-2016-01}. The first-level (L1) trigger is implemented in
hardware and uses a subset of detector information to reduce the event
rate to $\lesssim$\SI{100}{\kHz}. The subsequent, software-based
high-level trigger (HLT) selects events for recording.

\section{Datasets, event and track selection \label{sec:Datasets_trigger_and_tracking_cuts}}
The analysis presented in this \papertype\ uses a
$\sqs=8$~\TeV\ \pp\ dataset with an integrated luminosity of
19.4~\ifb\ obtained by the ATLAS experiment in 2012 and a
$\sqs=13$~\TeV\ \pp\ dataset recorded in 2015 and 2016 with
integrated luminosities of 3.2~\ifb\ and 32.9~\ifb, respectively. All
data used in the analysis come from data-taking periods
where the beam and detector operations were stable,
and the detector subsystems relevant for this analysis were fully operational.

The primary dataset used for the measurement was collected using  the
  dimuon or high-\pt single-muon triggers.
The primary triggers used in this analysis apply a combination of
  L1 and HLT muon-trigger algorithms~\cite{TRIG-2012-03,TRIG-2016-01} to select
  events with muons.
For the 8~\TeV\ analysis, events are selected using a single-muon trigger
  requiring $\pT > 36$~\GeV\ or a dimuon trigger requiring $\pT > 18$~\GeV\
  for the first muon and $\pT>8$~\GeV\  for the other.
For the 13~\TeV\ analysis a single-muon trigger with a \pt threshold of 24~\GeV\
  or a dimuon trigger with a \pt\ threshold of 14~\GeV\ for both muons
  are used to select events.
These triggers are complemented by other triggers depending on the running
  conditions over the course of the data taking.
A separate `zero bias' trigger is used to select events effectively at random
  but with the same luminosity profile as the muon triggers.
The zero-bias events are used to study charged-particle
  backgrounds arising from pile-up.
 Muons are reconstructed as combined tracks spanning both the
  ID and the MS~\cite{PERF-2014-05,PERF-2015-10}.
 For this analysis, muons associated with the event primary
 vertex~\cite{Aaboud:2016rmg} are selected and required to have $\pT > 20$~\GeV\ and $|\eta|<2.4$.
 Track quality requirements are imposed in both the ID and MS to
 suppress backgrounds.
In the analysis of the 13~\TeV\ data, muons are also
isolated using track-based and calorimeter-based isolation criteria studied in Ref.~\cite{PERF-2015-10}.
Events having exactly two such muons with opposite charge and
pair invariant mass between 80~\GeV\ and 100~\GeV\ are
  considered to be \Zboson-boson candidate events.
Data sample parameters are summarized in Table~\ref{tab:List-of-statistics}.
\begin{table}[htb!]
\begin{centering}
\caption{
The total integrated luminosity and number of \Zboson-tagged events in the
  datasets used in this analysis.
\label{tab:List-of-statistics}}
\begin{tabular}{c|S[table-format=4.1]S[table-format=4.1]c}
Year  & \multicolumn{1}{c}{\sqs [\TeV]} & \multicolumn{1}{c}{Luminosity [\ifb]}& \multicolumn{1}{c}{Number of events} \tabularnewline
\hline
2012         & 8         	& 19.4       	& $6.1\times10^6$   \tabularnewline
2015         & 13        	& 3.2        	& $1.6\times10^6$   \tabularnewline
2016         & 13         	& 32.9      	& $1.7\times10^7$   \tabularnewline
\end{tabular}
\par\end{centering}
\end{table}

All events considered in this analysis are required to have at least one
  reconstructed primary vertex with at least two associated tracks~\cite{Aaboud:2016rmg}.
Charged-particle tracks are reconstructed in the ID using the
  methods described in Refs.~\cite{STDM-2010-01,ATL-PHYS-PUB-2015-018}.
Tracks selected for this analysis are required to pass a set of quality requirements
  on the number of used and missing hits in the detector layers according to the
  track reconstruction model~\cite{ATL-PHYS-PUB-2015-018} and to have $\pt>0.4$~\GeV and $|\eta|<2.5$.
The ID tracks produced by \Zboson-boson decay muons are not
  included in the 2PC analysis.

The track reconstruction efficiencies, $\epsilon(\pt,\eta)$, are calculated 
  as a function of \pt\ and $\eta$ from Monte Carlo (MC) simulations
  of \pp\ collisions which are processed with a \textsc{Geant4}-based
	MC simulation~\cite{Agostinelli:2002hh} of the ATLAS detector~\cite{SOFT-2010-01}.
In the 8~\TeV\ data, the reconstruction efficiency ranges from
  approximately 70\% at $\pt=0.4$~\GeV\ to 80\% at $\pt=5$~\GeV\ for
  tracks at mid-rapidity ($|\eta|<0.5$).
The efficiency at forward rapidity ($2.0<|\eta|<2.5$)  varies between
  55\% at $\pt=0.4$~\GeV\ to 75\% at $\pt=5$~\GeV.
The 13~\TeV\ data were reconstructed with the IBL
  installed and this leads to a higher efficiency of 85\% (75\%) for
  mid-rapidity (forward) tracks.
The 13~\TeV\ efficiency shows only a very weak \pt\ dependence.

Tracks resulting from secondary particles and
tracks produced in pile-up interactions are suppressed by requiring:
\begin{gather}
|d_{0}| < 1.5~\mathrm{mm},  \,\,\,  |\omega|<0.75~\mathrm{mm}, \nonumber \\
\omega\equiv\left(z_{0}-\zvtx\right)\sin{\theta},
\label{eq:pile-up}
\end{gather}
where $d_0$ is the distance of the closest approach of the track
to the beam line in the transverse plane, $z_0$ and \zvtx are the
longitudinal coordinates of the track at $d_0$ and the \Zboson-tagged
collision vertex, respectively, and $\theta$ is the polar angle of the track.

\section{Two-particle correlations}\label{sec:2PC}
The study of two-particle correlations in this \papertype\ follows
   previous ATLAS measurements in \pp~collisions~\cite{HION-2015-09,HION-2016-01} ,
   with the additional complication of handling the pile-up, which is discussed later in Section\,\ref{sec:pile-up}.
The two-particle correlations are measured as a function of the relative
  azimuthal angle $\dphi\equiv\phi^a-\phi^b$ for particles separated by $|\deta|>2$.
This pseudorapidity gap is used to study the long-range component of the
  correlations~\cite{HION-2015-09,HION-2016-01}.
The labels  $a$ and $b$ denote the two particles in the pair, and in this \papertype\ are
  referred to as the `reference' and `associated' particles, respectively.
The correlation function is defined as:
\begin{eqnarray}
\label{eq:ana0}
C(\dphi) =\frac{S(\dphi)}{B(\dphi)}\;,
\end{eqnarray}
where $S$ represents the pair distribution constructed using all particle pairs that
  can be formed from tracks that are associated with the event containing the \Zboson-boson candidate
 and pass the selection requirements.
The $S$ distribution contains both the physical correlations between particle pairs and
  correlations arising from detector acceptance effects.
The pair-acceptance distribution $B(\dphi)$, is similarly constructed by choosing
  the two particles in the pair from different events.
The $B$ distribution does not contain physical correlations, but has detector acceptance
  effects in $\dphi$ identical to those in $S$.
By taking the ratio, $S/B$ in Eq.\,\eqref{eq:ana0}, the detector acceptance effects
  cancel out, and the resulting \ctwophi contains physical correlations only.
To correct $S(\dphi)$ and $B(\dphi)$ for the individual $\phi$-averaged inefficiencies
  of particles $a$ and $b$, the pairs are weighted by the inverse product of their
  tracking efficiencies $1/(\epsilon_a\epsilon_b)$.
Statistical uncertainties are calculated for \ctwophi\ using standard
  uncertainty propagation procedures with the statistical variance of
  $S$ and $B$ in each \dphi\ bin taken to be $\sum 1/(\epsilon_a\epsilon_b)^2$,
  where the sum runs over all of the pairs included in the bin.
Since the role of the reference and associated particles in the 2PC are different,
  when the reference and associated particles are from overlapping \pt\ ranges,
  the two pairings $a$--$b$ and $b$--$a$ are considered distinct and included separately
  in the pair distributions.
However, including both pairings correlates the statistical fluctuations at
  $\dphi=\phi^a-\phi^b$  and $\dphi=\phi^b-\phi^a$.
Thus the statistical uncertainties in the measured pair distributions are calculated
  by accounting for this correlation.
This is done by increasing the contribution to the statistical error
  in the $S$ and $B$ distributions for such correlated pairs by $\sqrt{2}$.
The two-particle correlations are used only to study the shape
  of the correlations in \dphi, and their overall normalization does not matter.
In this \papertype, the normalization of \ctwophi\ is chosen such that the
  \dphi-averaged value of \ctwophi\ is unity.

The strength of the long-range correlation can be quantified by
  extracting Fourier moments of the 2PC.
The Fourier coefficients of the 2PC are denoted \vnn and defined by:
\begin{eqnarray}
C(\dphi)=C_{0}(1+2\sum_{n}\vnn\cos(n\dphi)).
\label{eq:fourier_fit}
\end{eqnarray}
The \vnn are directly related to the single-particle anisotropies \vn described in
  Eq.\,\eqref{eq1}.
In the case where the \vnn\ entirely result from the  convolution of the single particle anisotropies,
  for reference and associated particles with $\pt=\pta$ and \ptb respectively,
  the $\vnn(\pta,\ptb)$ is the product of the $\vn(\pta)$ and $\vn(\ptb)$~\cite{HION-2011-01}, i.e.:
\begin{eqnarray}
v_{n,n}(\pta,\ptb)=v_{n}(\pta)v_{n}(\ptb).
\label{eq2}
\end{eqnarray}
Thus, the $\vn(\pta)$ can be obtained as:
\begin{eqnarray}
\vn(\pta)=\frac{\vnn(\pta,\ptb)}{\vn(\ptb)}=\frac{\vnn(\pta,\ptb)}{\sqrt{\vnn(\ptb,\ptb)}},
\label{eq:factorize}
\end{eqnarray}
where $\vnn(\ptb,\ptb)$ is the Fourier coefficient of the 2PC when
 both reference and associated particles are from the same \pt range.
This technique has been used extensively in heavy-ion collisions
  to obtain the flow harmonics~\cite{HION-2011-01}.
However, in \pp\ collisions a significant contribution to the 2PC arises from
  back-to-back dijets, which can correlate particles at large $|\deta|$.
These correlations must be removed before Eq.\,\eqref{eq2} or Eq.\,\eqref{eq:factorize}
 can be used.
In order to estimate the contribution from back-to-back dijets
  and other processes which correlate only a subset of all particles
  in the event, a template-fitting method was developed and used in two recent
  ATLAS measurements~\cite{HION-2015-09,HION-2016-01}.
The template-fitting procedure assumes that:
(1) the jet--jet correlation has the same shape in \dphi in
      low-multiplicity and in higher-multiplicity events;
      the only change is in the relative contribution of the dijets to the 2PC,
(2) at low-multiplicity most of the structure of the 2PC arises from
      back-to-back dijets, i.e. the shape of the dijet correlation can be
      obtained from low-multiplicity events.
With the above assumptions, the correlation in higher-multiplicity
  events $C(\dphi)$, is then described by a template fit, $\ctempl(\dphi)$
  consisting of two components:
 1) the correlation that accounts for the dijet contribution, $\cperi(\dphi)$,
   measured in low-multiplicity events and scaled by a factor $F$, and
 2) a long-range harmonic modulation, $\cridge(\dphi)$:

\begin{eqnarray}
\ctempl(\dphi) & =F\cperi(\dphi)+G\left(1+2\sum_{n=2}v_{n,n}\cos(n\dphi)\right)\label{eq:template_new}\\
 & \equiv F\cperi(\dphi)+\cridge(\dphi),
\label{eq:template}
\end{eqnarray}

where the coefficient $F$ and the \vnn are fit parameters
  adjusted to reproduce the \ctwophi.
The coefficient $G$ is not a free parameter, but is fixed by the
  requirement that the integrals of the $\ctempl(\dphi)$ and $C(\dphi)$
  over the full $\dphi$ range are equal.

In this analysis, the  $\cperi(\dphi)$ is obtained for the
  20--30 multiplicity interval, where the multiplicity is evaluated
  using tracks satisfying the selection criteria described in
  Section~\ref{sec:Datasets_trigger_and_tracking_cuts} and corrected for pile-up
  as described below in Section~\ref{sec:pile-up}.
This choice of peripheral reference is different from the analysis in
  Refs.\,\cite{HION-2015-09,HION-2016-01} where the 0--20 interval
  was used.
This change is due to the relative rarity of events having less than
  20 tracks  in the \Zboson-tagged sample, which would impair the
  statistical precision of the peripheral reference.
The systematic uncertainty associated with choosing a higher-multiplicity
 peripheral reference is evaluated by comparing \vtwo
  results obtained when using other peripheral intervals,
  including the 0--20 track multiplicity interval.

\section{Pile-up subtraction}
\label{sec:pile-up}

Selected events from  all three data-taking periods contain significant pile-up,
which has a direct impact on the measurement of the two-particle correlations.
The tracks used in the  analysis, selected using the requirements in  Eq.~\eqref{eq:pile-up},
are associated with the collision vertex that includes the \Zboson boson.
The residual contribution from
pile-up tracks to the measured distributions is evaluated and corrected on a
statistical basis. The correction procedure, based on an event mixing technique,
is explained in this section. The main parameters affecting the pile-up are described below.
Track categories used in the analysis are introduced in Section~\ref{sec:pile}, and a description of the event
mixing technique and its performance can be found in Section~\ref{sec:mixed}.
Section~\ref{sec:estimator} introduces the parameter \avgntrk, the average number of
pile-up tracks expected in the event. The parameter \avgntrk fully defines properties of the
residual pile-up as discussed in Section~\ref{sec:properties} and therefore can be used to
correct the measured multiplicity as explained in Section~\ref{sec:unfold}.
Section~\ref{sec:pair-correction} derives the algorithm in which the additional
event sample obtained with the mixing procedure is used in the measurement of
the two-particle correlations.

The two main time-dependent characteristics which primarily define the
  pile-up contributions to the measured events are the distribution of
  the \Zboson-boson interaction longitudinal vertex position, \zvtx,
  and the instantaneous luminosity which is characterized by the per-crossing
  number of collisions, \nint.
Distributions of \zvtx\ and \nint\ are shown in panels (a) and (b) of Figure~\ref{fig:mains}, respectively, for the three data-taking periods
  used in the measurement.
The mean values of the \zvtx\ distributions are close to the centre of the
  ATLAS detector and are slightly negative.
The RMS of the \zvtx\ distributions vary period by period from approximately
 48~mm to 35~mm.
The instantaneous luminosity conditions yield
  an average number of interactions per bunch crossing $\langle\nint\rangle\approx20$,
  15 and 26 in the years 2012, 2015 and 2016, respectively.

\begin{figure*}[!htb]
\begin{center}
\includegraphics[width=0.99\textwidth]{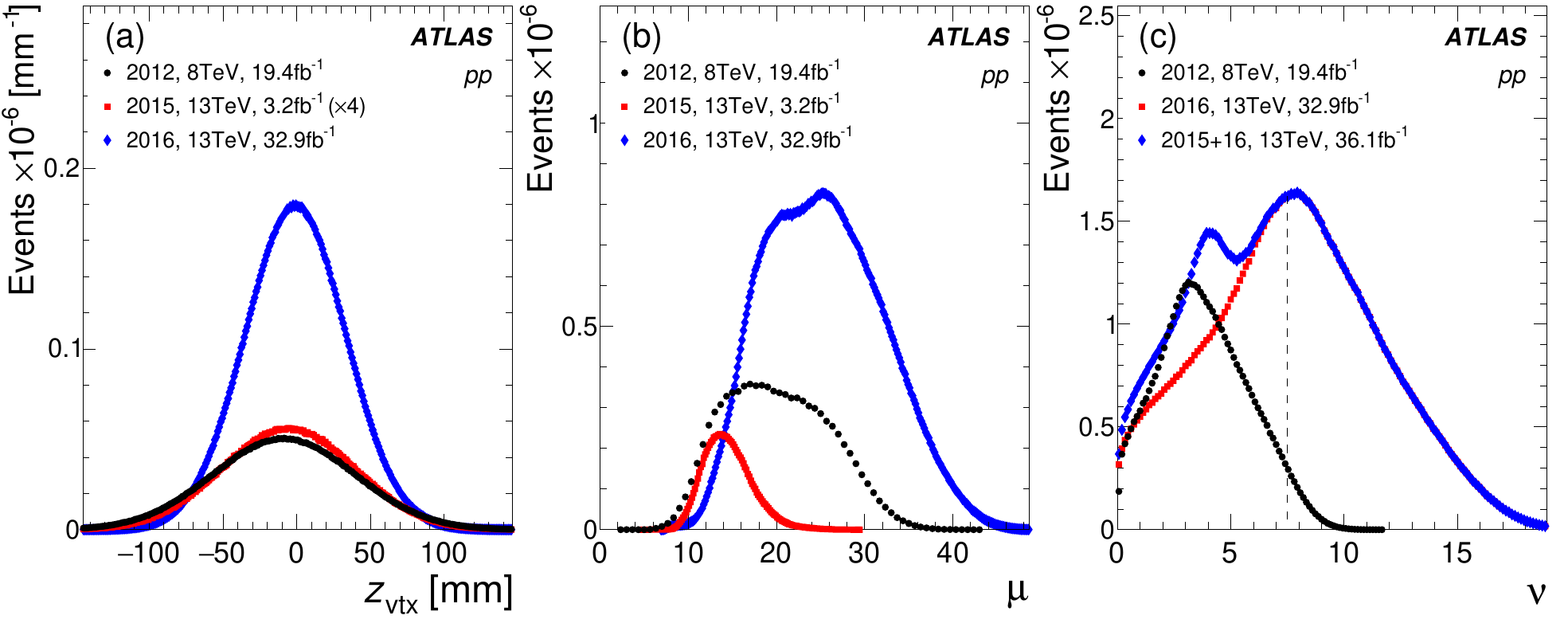}
\caption{
Distribution of parameters:
  (a) vertex position \zvtx,
  (b) instantaneous luminosity parameter measured as the number of interactions per bunch crossing $\mu$,
  (c) the average number of pile-up tracks accepted in the analysis \avgntrk,
in the three data-taking periods.
The vertical dashed line in the right plot at $\avgntrk=7.5$ indicates the criterion below which
  the events are selected for the analysis.
}
\label{fig:mains}
\end{center}
\end{figure*}

The \zvtx\ position and the instantaneous luminosity define the parameter
  that is used to characterize pile-up in the analysis.
This parameter, denoted \avgntrk, is the average number of background tracks
  per event from pile-up interactions that enter the analysis.
Its distribution is shown in panel (c) of Figure~\ref{fig:mains} and derivation is given in Section~\ref{sec:mixed}. 
The mean values of \avgntrk over the datasets, denoted by $\langle\avgntrk\rangle$, are
  about 4 in the $\sqs=8$~\TeV\ data and above 7 in the $\sqs=13$~\TeV\ data.
The 2015 data sample is only 10\% as large as the 2016 sample, but
the pile-up condition $\langle\avgntrk\rangle$ in this sample is less than half as large. The 2015 contribution forms the lower peak in the distribution of \avgntrk shown in panel (c) of Figure~\ref{fig:mains}.

\subsection{Event categories}
\label{sec:pile}
Tracks and track pairs that pass the selections described in Section~\ref{sec:Datasets_trigger_and_tracking_cuts}  and belong to a single event are referred to as \textit{Direct}. The \Direct contribution consists of tracks and pairs arising from the same interaction as the \Zboson boson --- referred to as \textit{Signal} --- and from pile-up interactions --- referred to as \textit{Background}.
The presence of the \Bkg contribution in the \Direct data affects both the
  number of measured tracks (\ntrk) and the two-particle correlations.
To extract the \Signal, the contribution of the \Bkg to the \Direct data needs
  to be subtracted.
For this purpose, a sample of events --- referred to as \textit{Mixed} and, ideally,
  equivalent to the \Bkg events --- is constructed using a random selection procedure.
In the following sections, the numbers of tracks in the different event categories
  are denoted by \ntrkdirect, \ntrksignal, \ntrkbackground and \ntrkmixed.

\subsection{\Mixed event sample}
\label{sec:mixed}

The \Mixed event sample is constructed using a random selection procedure
which is an extension of a technique used in Ref.~\cite{STDM-2011-42}.
It constructs an event that is similar to the \Direct event, but contains
no \Signal component.
It is done by requiring the longitudinal impact parameter of the track in
one event to be within 0.75~mm of the \zvtx measured in another
event (Eq.~\eqref{eq:pile-up}) taken during the same beam fill of the LHC.
To account for differences between \zvtx distributions from different LHC fills during the data taking, the analysis uses reduced values of \nint and \zvtx that are:
\begin{eqnarray}
(\rint,\rvtx)=\biggl(\frac{\nint}{\sqrt{2\pi}\,\mathrm{RMS}(\zvtx)},\frac{\zvtx-\langle\zvtx\rangle}{\mathrm{RMS}(\zvtx)}\biggr),
\label{eq:reduced}
\end{eqnarray}
where $\langle\zvtx\rangle$ and $\mathrm{RMS}(\zvtx)$ are the mean and
width of the \zvtx distribution parameterized as a function of time during the data taking,
and $\sqrt{2\pi}$ comes from the normalization of a Gaussian probability distribution.
\Direct and \Mixed events are required to have \rint values within 0.01 mm$^{-1}$ of each other, a parameter chosen in the analysis to be small enough to ensure the same instantaneous luminosity condition for both events.

Two event samples can be used by the random selection procedure to construct
  \Mixed events,  one obtained with a random trigger (zero bias sample),
  and the other obtained with  the same trigger as the \Direct event sample.
In the latter case, an additional condition must be used that requires the
  distance between the \zvtx positions in two events to be $|\Delta\zvtx|>15$~mm.
This is to ensure that the interaction that triggered the
  event recording and has particle counts and kinematics different from the
  inclusive (pile-up) interactions does not contribute to the \Mixed event which
  aims to reproduce only the \Bkg component.
\Mixed events constructed with both samples yield identical results, so
  the analysis uses the data sample with \Zboson bosons, which automatically
  ensures identical data-taking conditions in \Direct and \Mixed events.
The procedure is validated using a MC simulation sample where \Zboson-boson events 
  from 8~\TeV\ \pp collisions are generated with the \SHERPA\
  event-generator~\cite{Gleisberg:2008ta} and reconstructed with pile-up conditions
  corresponding to the 8~\TeV\ dataset used in this analysis.
This pile-up is simulated using \PYTHIA v8.165~\cite{Sjostrand:2007gs} with parameter values set
  according to the A2 tune~\cite{ATL-PHYS-PUB-2012-003} and the MSTW2008LO PDF set~\cite{Martin:2009iq}.
Implementing the procedue in the MC sample shows
  that the distributions found in \Mixed events are equivalent to those in
  the \Bkg events.
To suppress undesired statistical fluctuations in the \Mixed event sample the
  random selection procedure is performed 20 times for each \Direct event.

Figure~\ref{fig:pv_pointing} shows the average track density in \Direct
  and \Mixed events for different values of \rvtx and \rint.
The three panels correspond to three different intervals of \rvtx position
  and the different markers denote different \rint intervals.
For the distributions corresponding to \Direct events, the contribution from the
  \Signal tracks forms the peak at $\omega=0$, and the contribution from the
  \Bkg tracks produces a slowly changing distribution outside and under the peak.
The vertical axis in Figure~\ref{fig:pv_pointing} is restricted to low values in order to clearly show
  the contribution from \Mixed events, so the peaks at $\omega=0$ are truncated.
The solid lines are parabolic fits to the \Direct\ track distributions outside
  the peak regions and then interpolated under the peaks.
There is good agreement between the results of the fits and the results of
  \Mixed events in the region under the peak.
At values of $|\omega|>2.45$~mm, \Mixed curves in all $(\rint, \rvtx)$
  intervals depart from the \Direct ones.
This is due to the contribution from collisions that fired the trigger.
The \ntrk in them is larger than in the pile-up interaction and causes the excess.
However, due to the requirement that $|\Delta\zvtx|$ between the \Direct event
  and the event used by the random selection procedure must be greater than 15~mm,
  no tracks from triggered collisions can affect the region of $|\omega|<0.75$~mm,
  where agreement between the fitted \Direct and \Mixed events is good for the purpose of the analysis.

\begin{figure*}[!t]
\begin{center}
\includegraphics[width=0.99\textwidth]{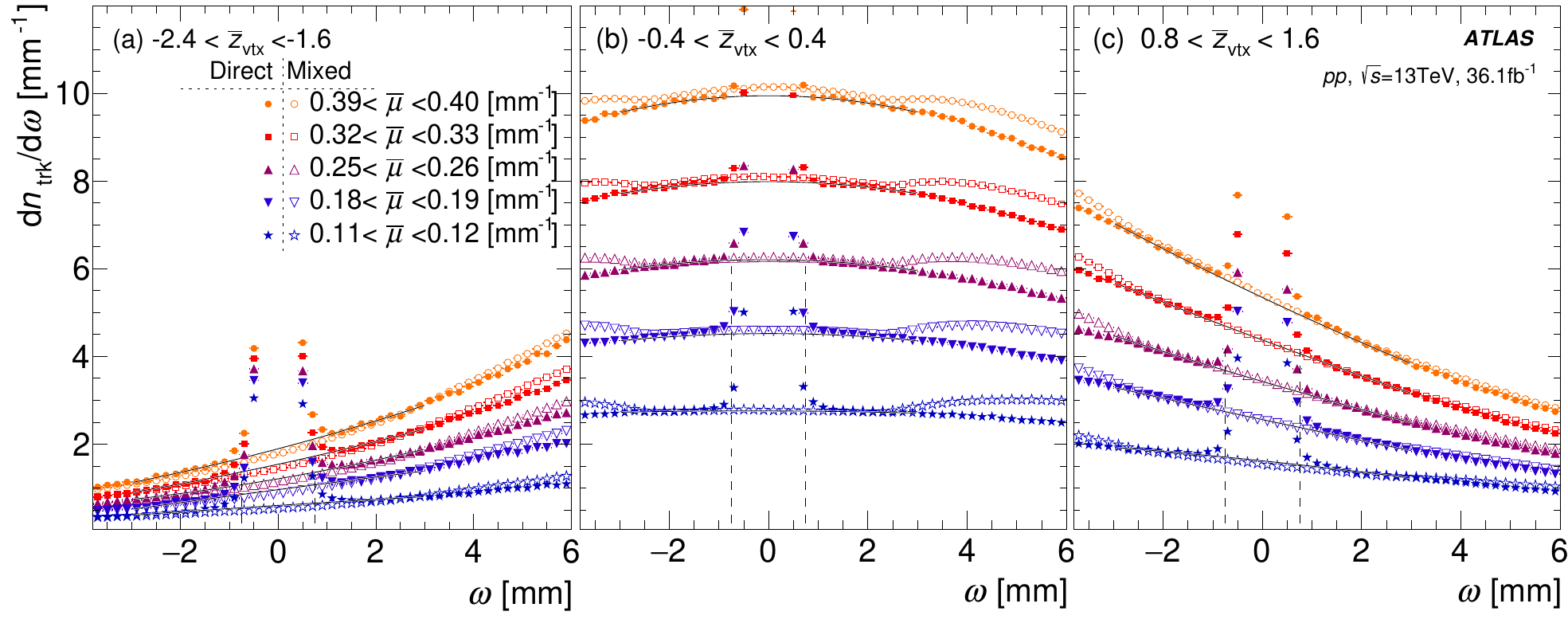}
\caption{
The number of tracks per mm as a function of $\omega$, defined by Eq.~\eqref{eq:pile-up},
  for \Direct (solid markers) and \Mixed events (open markers).
The three panels show results in different intervals of the reduced vertex
  \rvtx position and different marker colours correspond to several intervals
  of reduced \rint.
The solid lines are parabolic fits to \Mixed events in the region $|\omega|<3$~mm
and the vertical dashed
  lines show the acceptance window $|\omega|<0.75$~mm.
  The vertical axis is restricted to low values in order to show the \Mixed events, so the peaks at $\omega = 0$ are truncated.
}
\label{fig:pv_pointing}
\end{center}
\end{figure*}

Based on the level of agreement shown in Figure~\ref{fig:pv_pointing},
  and on the MC simulation studies, this analysis uses
  the approximation that the features of the \Mixed events
  (momentum, pseudorapidity distributions of tracks and two particle correlations)
  are equivalent to those of the \Bkg events.

\subsection{\Bkg estimator}
\label{sec:estimator}
The \Mixed track density under the peak ($|\omega|$<0.75~mm) shown in Figure~\ref{fig:pv_pointing} for
  \Mixed events is plotted in the left panel of Figure~\ref{fig:fit_results} as a
  function of \rint for different \rvtx.
\begin{figure*}[!htb]
\begin{center}
\includegraphics[width=0.8\textwidth]{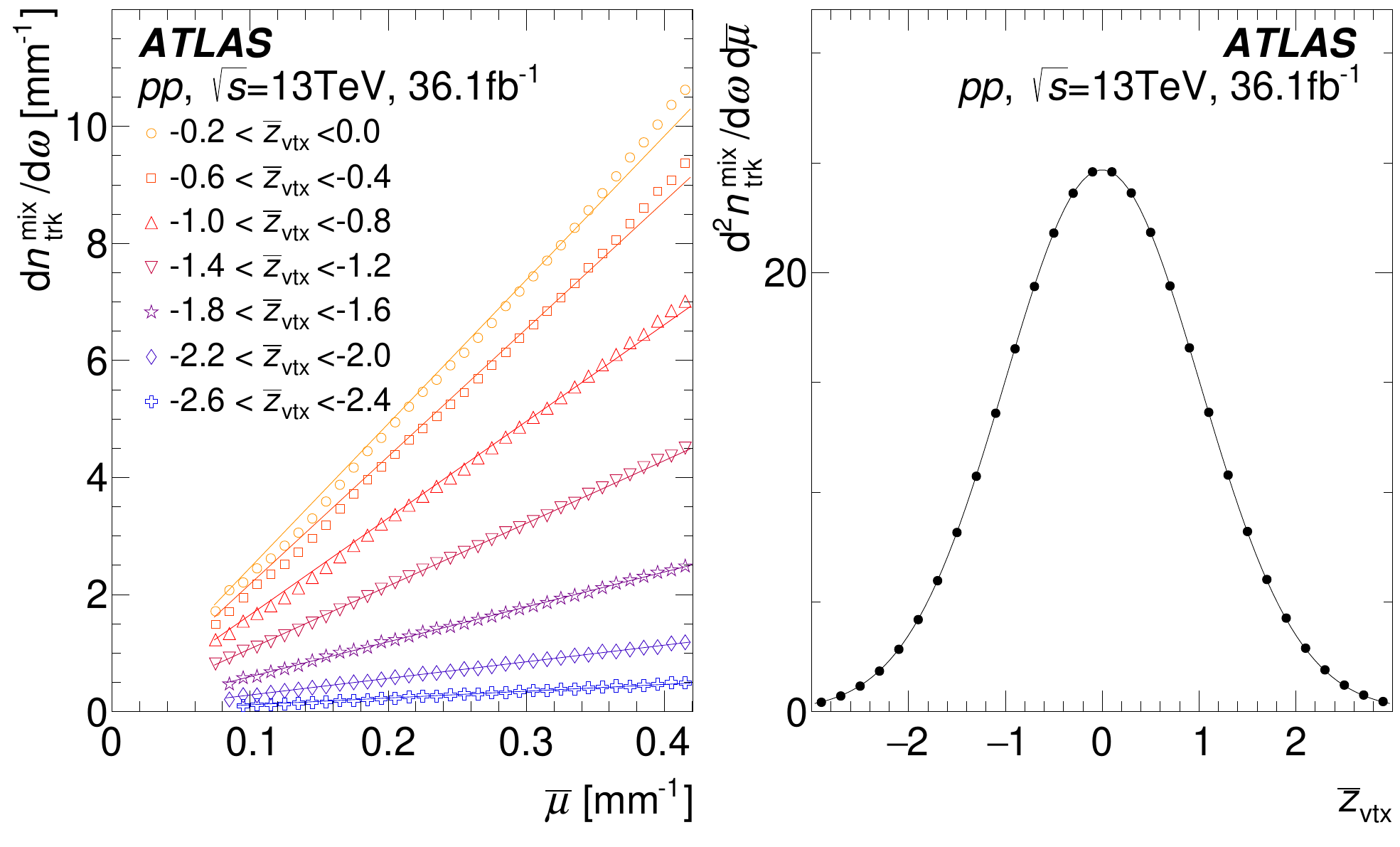}
\caption{
Left: The number of tracks in \Mixed events per mm at $\omega=0$ as a function of \rint. Different marker colours correspond to selected \rvtx intervals. Not all  intervals are shown for figure clarity. Solid lines are fits assuming scaling of track density with \rint. Right: Slopes of the lines shown in the left panel as a function of \rvtx fitted to a Gaussian shape.
}
\label{fig:fit_results}
\end{center}
\end{figure*}
Only intervals in $\rvtx<0$ are plotted since there is a symmetry around $\rvtx=0$.
The distribution of $\mathrm{d}\ntrkm/\mathrm{d}\omega$ evaluated as a function of
  \rint shows that track density is proportional to the interaction density:
  $\mathrm{d}\ntrkm/\mathrm{d}\omega\propto\rint$.
The proportionality coefficients, $\mathrm{d^{2}}\ntrkm/(\mathrm{d}\omega\mathrm{d}\rint)$,
  are determined by fitting a linear function to the $\mathrm{d}\ntrkm/\mathrm{d}\omega(\rint)$ distribution.
The small residual deviations from this linear fit are taken into account while estimating systematic uncertainties;
  they are primarily present in the regions of $(\rint,\rvtx)$ that are not used in the analysis.
The dependence of these coefficients on \rvtx is shown in the right panel
of Figure~\ref{fig:fit_results}. One can see that $\mathrm{d^{2}}\ntrkm/(\mathrm{d}\omega\mathrm{d}\rint)(\zvtx)$ is Gaussian with mean at zero and width very close to unity.
This is expected as the \rvtx, according to Eq.~\eqref{eq:reduced}, is already a reduced parameter.
Using the equivalence $\Bkg\equiv\Mixed$, the average number of \Bkg tracks can be expressed as:
\begin{eqnarray}
\avgntrk \equiv \big\langle n_{\mathrm{trk}}^{\mathrm{bkg}}\big\rangle = 2\omega_{0}\;\frac{\mathrm{d}^{2}\ntrkm}{\mathrm{d}\omega~\mathrm{d}\rint}\bigg|_{\rvtx=0}\mathrm{Gauss}(\rvtx) \rint,
\label{eq:bkg}
\end{eqnarray}
where $\omega_{0}=0.75$~mm is half of the width of the track acceptance window, $\mathrm{d}^{2}\ntrkm/(\mathrm{d}\omega \mathrm{d}\rint)|_{\rvtx=0}$ is the coefficient defined by particle production in inclusive \pp\ collisions and by the detector rapidity coverage and efficiency,
and $\mathrm{Gauss}(\rvtx)$ is a Gaussian function with mean equal to 0 and a variance of 1.0.

\subsection{Properties of \Mixed events}
\label{sec:properties}
The parameters \rint and \rvtx factorize in Eq.~\eqref{eq:bkg}. There is only a scaling coefficient between \avgntrk and the interaction density $\mathrm{Gauss}(\rvtx)\rint$, such that the same \avgntrk can be reached at low instantaneous luminosity and close to the centre of the \rvtx interval, or at high instantaneous luminosity and large \rvtx. Using the MC simulations and \Mixed events taken at different $(\rint, \rvtx)$ one can find that not only the average value, but also the shape of the \ntrkb distribution are the same for the same interaction density $\mathrm{Gauss}(\rvtx)\rint$ and consequently for the same \avgntrk.  Events are therefore fully characterized with respect to their background conditions by \avgntrk, calculated using Eq.~\eqref{eq:bkg}.
This is demonstrated in Figure~\ref{fig:nu_regions} for three intervals: $\avgntrk<0.5$, $3<\avgntrk<3.5$ and $7<\avgntrk<7.5$.
For each interval the probability distributions of \Mixed tracks
  $P_{\mathrm{mix}}$ obtained without any restriction on \rvtx, are compared with
  the $P_{\mathrm{mix}}$ distributions obtained when restricting the \rvtx\ to
  three different intervals of $|\rvtx|<0.2$,  $0.2<\rvtx<0.8$, and $0.8<\rvtx<3$.
Although no constraint is imposed on \rint, its value varies over a different range
  for each \rvtx interval to provide \avgntrk according to Eq.~\eqref{eq:bkg}.
  Some distributions are not shown because it is impossible to find low-\avgntrk conditions
  at the centre of the \zvtx distribution at any $\mu$ shown in Figure~\ref{fig:mains}.
  The upper panels of the figure show the $P_{\mathrm{mix}}$ distributions and
  the lower panels show the ratios of the $P_{\mathrm{mix}}$ distributions in
  each \rvtx interval to the $P_{\mathrm{mix}}$ distribution measured without
  any restriction on \rvtx.
 The ratios in the lower panels are consistent with unity within 5\% in most cases, demonstrating that
  for a given \avgntrk the shape of the $P_{\mathrm{mix}}$ distribution does not depend
  on \rvtx or \rint. Residual deviations are due to tracking efficiency variation along the
  beam axis, accuracy of determining $\mu$, and deviations from the parameterizations used in Eq.~\eqref{eq:bkg}.

\begin{figure*}[!htb]
\begin{center}
\includegraphics[width=0.99\textwidth]{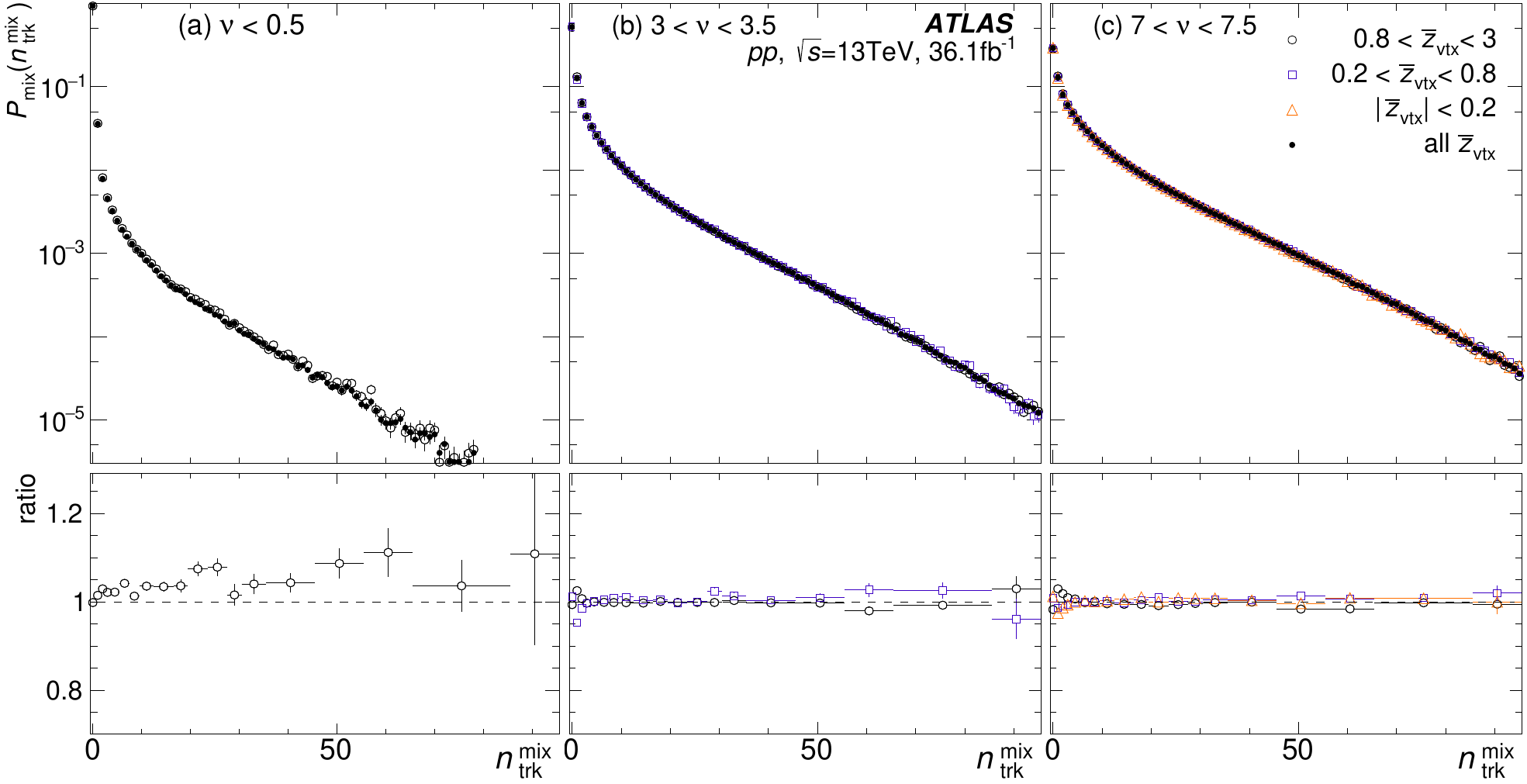}
\caption{
The upper panels show the probability distributions for the \ntrkm
  measured without any restriction on \rvtx (filled markers)
  as well as in three different \rvtx intervals (open markers).
From left to right, the panels correspond to \avgntrk ranges of (a) $\avgntrk<0.5$,
 (b) $3<\avgntrk<3.5$ and (c) $7<\avgntrk<7.5$.
The lower panels show the ratio of the $P_{\mathrm{mix}}$ in the
  \rvtx intervals to the $P_{\mathrm{mix}}$ distribution obtained without any restriction on \rvtx.
 Vertical bars are the statistical uncertainty.
}
\label{fig:nu_regions}
\end{center}
\end{figure*}

The probability distributions for the \ntrk found under different \avgntrk conditions
  are shown in Figure~\ref{fig:ntrk_shapes}.
The left and right panels display probabilities $P_{\mathrm{dir}}$ and $P_{\mathrm{mix}}$ for the \Direct and \Mixed events respectively.
The continuous lines are the fits to the data points to smooth the statistical
  fluctuations at high \ntrk.
\begin{figure*}[!htb]
\begin{center}
\includegraphics[width=0.8\textwidth]{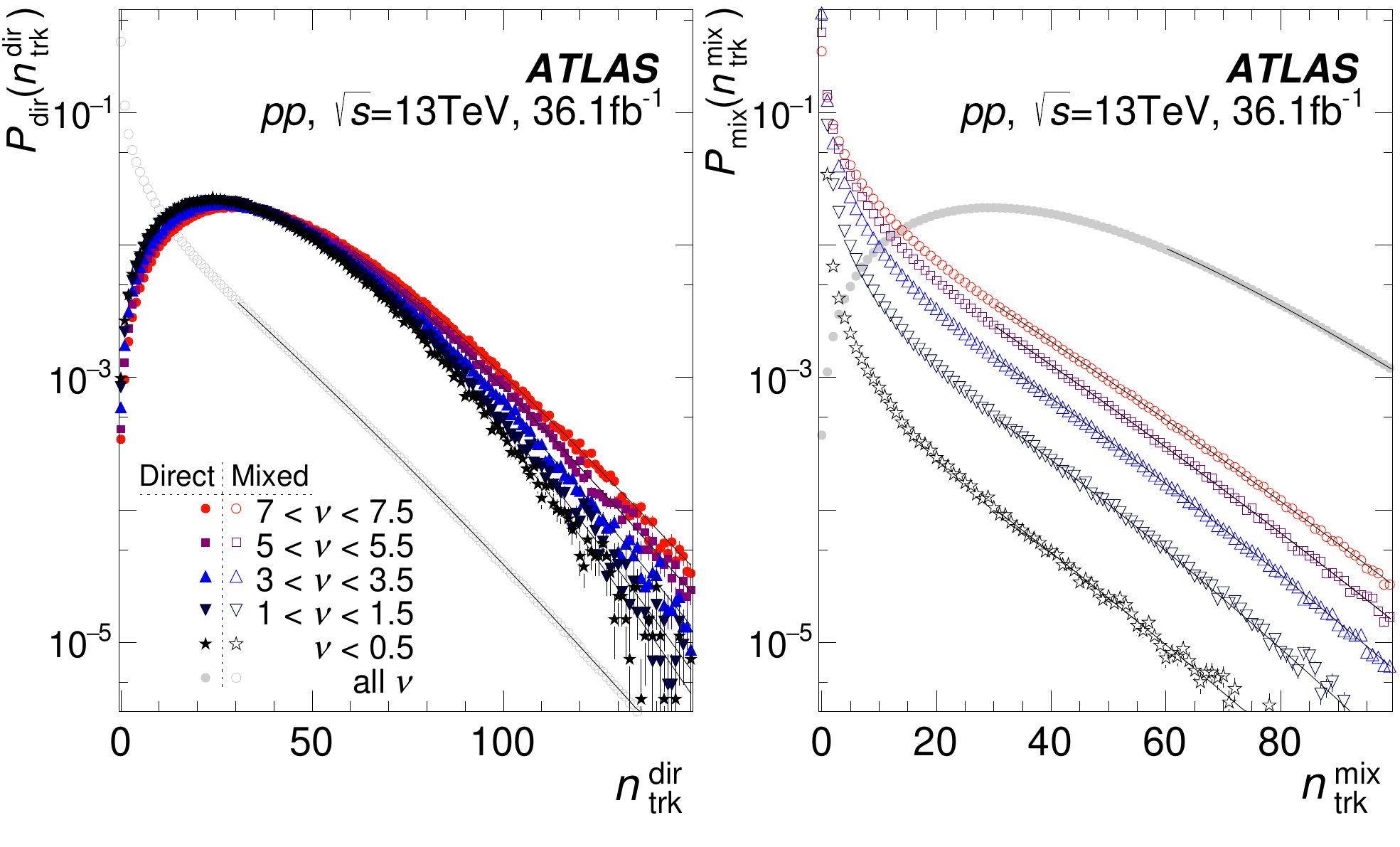}
\caption{
Probability distributions for the \ntrk in \Direct events (left) and
  \Mixed events (right).
The different coloured markers correspond to different values of \avgntrk.
The grey distribution in each panel indicates the distributions from the
  other panel averaged over the sample, and is shown for comparison.
The lines are fits to data points.
The $x$-axis ranges are different in the two panels.
}
\label{fig:ntrk_shapes}
\end{center}
\end{figure*}

Figure~\ref{fig:ntrk_shapes} shows that the \Bkg tracks affect \Direct distributions differently, depending on the \ntrk regions. Assuming that the lower \ntrkd distribution, shown with black markers $\big(\avgntrk<0.5\big)$, resembles the no pile-up condition, Figure~\ref{fig:ntrk_shapes} implies that at $\ntrkdirect>100$ the \Direct event distributions at high \avgntrk are dominated by the \Bkg tracks, rising by an order of magnitude relative to black markers for the highest \avgntrk measured in the event sample. Averaged over the sample, the distribution for \ntrkd is shown in the right panel and for \ntrkm in the left panel for comparison with the distributions of the opposite type. The mean numbers of tracks in those distributions are 30 and 4 respectively.

\subsection{Correction of \ntrk distribution}
\label{sec:unfold}

The \ntrks distributions are derived by unfolding the \ntrkd distributions.
Transition matrices required for that are constructed from the data.
For the analysis of the 2PC the same matrices are used to remap the
  correlation coefficients measured for \ntrkd to \ntrks explained later in Section~\ref{sec:pair-correction}.
These matrices  are constructed from the data using
  the distributions shown in Figure~\ref{fig:ntrk_shapes}:
\begin{eqnarray*}
M\big(\avgntrk,\ntrks,\ntrkd\big)=P_{\mathrm{dir}}\big(\avgntrk<0.5,\ntrks\big)~P_{\mathrm{mix}}\big(\avgntrk,\ntrkd-\ntrks\big).
\label{eq:matrix}
\end{eqnarray*}
The matrices are calculated using the \ntrkd distribution measured at the lowest \avgntrk ($\avgntrk<0.5$) as a proxy for the \ntrks distribution and \ntrkm distributions corresponding to different intervals of \avgntrk.
The probabilities to find \ntrkd shown in the left panel of Figure~\ref{fig:ntrk_shapes} are multiplied by the probabilities to find \ntrkm, shown in the right panel of Figure~\ref{fig:ntrk_shapes}. The product of the two probabilities is the matrix element for $\big(\ntrks,\ntrkd\big)$ using the relation $\ntrks=\ntrkd-\ntrkm$.
For high numbers of tracks, the fits shown in Figure~\ref{fig:ntrk_shapes} are used to
  suppress statistical fluctuations.
Examples of the transition matrices for two different \avgntrk are shown in Figure~\ref{fig:matrix}.

\begin{figure*}[!htb]
\begin{center}
\includegraphics[width=0.99\textwidth]{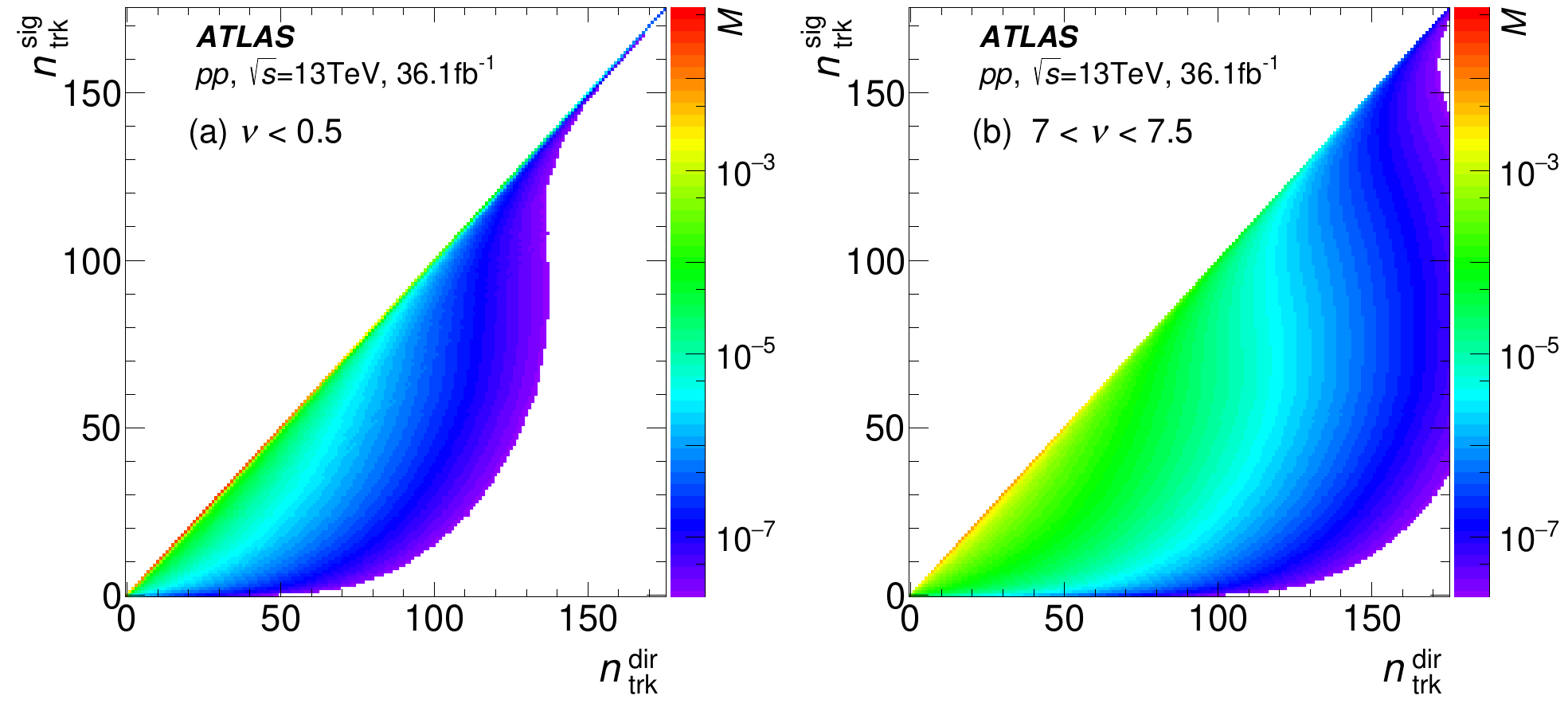}
\caption{
Data-driven transition matrices corresponding to intervals (a) $\avgntrk<0.5$ and (b) $7<\avgntrk<7.5$ that are used for remapping.
}
\label{fig:matrix}
\end{center}
\end{figure*}

The contour lines of the matrices have a distinct `spinnaker' shape with the amount of `drag' increasing with \avgntrk. At high \avgntrk, the higher values of \ntrk in \Direct events become only weakly correlated with the \ntrk in \Signal events. The right panel of Figure~\ref{fig:matrix} shows that the largest number of tracks in \Direct events corresponds to relatively moderate \Signal \ntrk smeared by the \Bkg. This effect limits the range of \ntrk values where the pile-up data samples can be analysed, and the limit depends on the value of \avgntrk.

Each \Direct event with a given \ntrk contains contributions from \Signal events
  with any number of tracks such that $\ntrks\le\ntrkd$.
Those contributions are calculated from the transition matrices,
  shown in Figure~\ref{fig:matrix}, by making a projection of \ntrkd
  onto \ntrks for a given value of \ntrkd.
These projections are shown in Figure~\ref{fig:refold} for two intervals of \avgntrk.

\begin{figure*}[!htb]
\begin{center}
\includegraphics[width=0.85\textwidth]{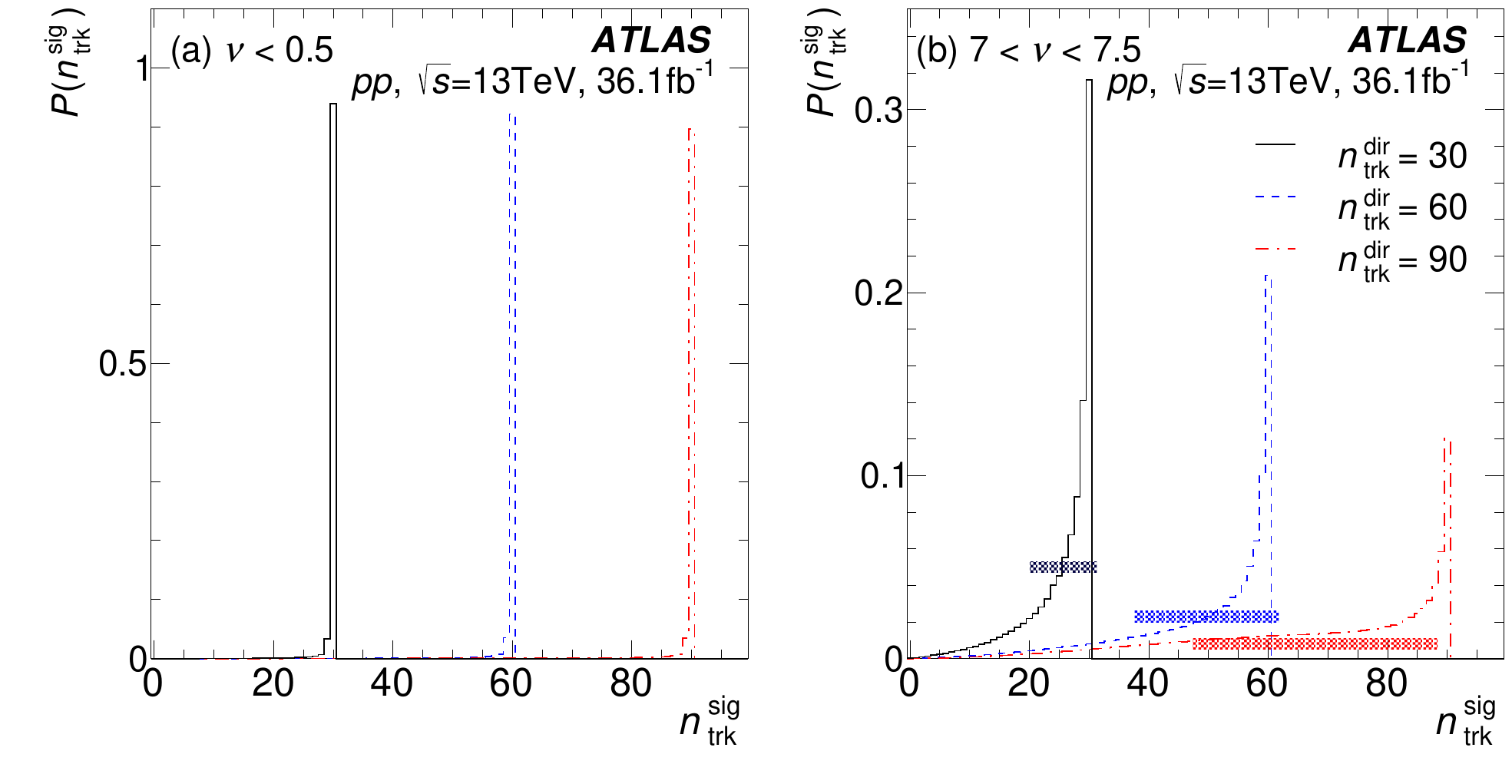}
\caption{
The probability of a \Signal event with multiplicity \ntrks,
  to contribute to a \Direct event with $\ntrkd=30$, 60 and 90
  (solid, dashed, and dotted-dashed), as a function of \ntrks.
The shaded bands denote the horizontal range equal to the mean~$\pm$~RMS value of
  the histogram with the corresponding colour.
Panel (a) is for $\avgntrk < 0.5$ and panel (b) for $7 < \avgntrk < 7.5$.
}
\label{fig:refold}
\end{center}
\end{figure*}

The histograms in Figure~\ref{fig:refold} are examples of probability distributions
  of the \ntrks contributing to \Direct events with  $\ntrkd=30,\,60$, and 90.
At low \avgntrk, shown in the left panel, the distributions are
  narrow and peaked at $\ntrks=\ntrkd$.
For this low pile-up condition, more than 85\% of \Direct events do not have
  even one \Bkg track.
The situation is different for high \avgntrk (right panel) where the
  contributions to \Direct events come from a wide range of \Signal events
	with smaller \ntrk.
The shaded bands shown in the plot are centred horizontally at the mean values
  of \ntrks contributing to the \Direct events, and have widths equal to
	2$\times$RMS of the corresponding distributions.
Distributions for high values of \ntrkd become increasingly wider as shown in
  the right panel of Figure~\ref{fig:refold}.
This figure demonstrates that with increasing \avgntrk it becomes impossible to
  accurately determine to what \ntrks the measurement belongs.
Figure~\ref{fig:refold} shows that the presence of pile-up degrades the
  resolution with which one can measure \ntrks.
As described in Section~\ref{sec:pair-correction}, this analysis is restricted to $\avgntrk<7.5$ because otherwise the pile-up is too large to correct the two particle correlations.

\newcommand{\probsig }{\ensuremath{P\big(\ntrksignal\vert\avgntrk,\ntrkdirect\big)}}

\newcommand{\dnpair     }{\ensuremath \mathrm{d}N_{\mathrm{pair}}}
\newcommand{\dnpairsig  }{\ensuremath \mathrm{d}N_{\mathrm{pair}}^{\mathrm{sig}}}
\newcommand{\dnpairbkg  }{\ensuremath \mathrm{d}N_{\mathrm{pair}}^{\mathrm{bkg}}}
\newcommand{\dnpairdir  }{\ensuremath \mathrm{d}N_{\mathrm{pair}}^{\mathrm{dir}}}
\newcommand{\dnpairmix  }{\ensuremath \mathrm{d}N_{\mathrm{pair}}^{\mathrm{mix}}}

\newcommand{\dnsing     }{\ensuremath \mathrm{d}N_{\mathrm{trk}}}
\newcommand{\dnsingsig  }{\ensuremath \mathrm{d}N_{\mathrm{trk }}^{\mathrm{sig}}}
\newcommand{\dnsingbkg  }{\ensuremath \mathrm{d}N_{\mathrm{trk }}^{\mathrm{bkg}}}
\newcommand{\dnsingdir  }{\ensuremath \mathrm{d}N_{\mathrm{trk }}^{\mathrm{dir}}}
\newcommand{\dnsingmix  }{\ensuremath \mathrm{d}N_{\mathrm{trk }}^{\mathrm{mix}}}

\newcommand{\ddphi   }{\ensuremath{\mathrm{d}\dphi}}
\newcommand{\ddphiab }{\ensuremath{\mathrm{d}\Delta\phi^{ab}}}
\newcommand{\ddphisig}{\ensuremath{\mathrm{d}\dphi}} 
\newcommand{\ddphibkg}{\ensuremath{\mathrm{d}\dphi}} 
\newcommand{\ddphidir}{\ensuremath{\mathrm{d}\dphi}} 
\newcommand{\ddphimix}{\ensuremath{\mathrm{d}\dphi}} 
\newcommand{\ddelta  }{\ensuremath{\delta(\Delta\phi-\Delta\phi^{ab})}}

\newcommand{\dphia   }{\ensuremath{\mathrm{d}\phi^{a}}} 
\newcommand{\dphib   }{\ensuremath{\mathrm{d}\phi^{b}}} 
\newcommand{\dphibkga}{\ensuremath{\mathrm{d}\phi^{a}}} 
\newcommand{\dphibkgb}{\ensuremath{\mathrm{d}\phi^{b}}} 
\newcommand{\dphisiga}{\ensuremath{\mathrm{d}\phi^{a}}} 
\newcommand{\dphisigb}{\ensuremath{\mathrm{d}\phi^{b}}} 
\newcommand{\dphidira}{\ensuremath{\mathrm{d}\phi^{a}}} 
\newcommand{\dphidirb}{\ensuremath{\mathrm{d}\phi^{b}}} 
\newcommand{\dphimixa}{\ensuremath{\mathrm{d}\phi^{a}}} 
\newcommand{\dphimixb}{\ensuremath{\mathrm{d}\phi^{b}}} 

\newcommand{\sal}{\ensuremath{\bigg\langle}} 
\newcommand{\sar}{\ensuremath{\bigg\rangle}} 
\newcommand{\dal}{\ensuremath{\Bigg\langle\!\bigg\langle}} 
\newcommand{\dar}{\ensuremath{\bigg\rangle\!\Bigg\rangle}}  
\newcommand{\dals}[1]{\ensuremath{\Bigg\langle\ #1\bigg\langle}} 
\newcommand{\dars}[1]{\ensuremath{\bigg\rangle\ #1\Bigg\rangle}}  

\newcommand{\nevtdir}{\ensuremath n_{\mathrm{evt}}^{\mathrm{dir}}}

\subsection{Correction for the pair-distribution}
\label{sec:pair-correction}
This section describes the pile-up correction procedure for the pairs 
  that are obtained by correlating particle pairs in \Direct\ events. 
The \dphi\ distribution of track-pairs found in one \Direct\ event 
  can be formally written as: 
\begin{eqnarray}
\label{eq:corr_long}
\frac{\dnpairdir}{\ddphi}
&=& \sum_{a}^{\ntrkdirect}    \sum_{b\neq a}^{\ntrkdirect}     \ddelta  \nonumber \\  
&=& \sum_{a}^{\ntrksignal}    \sum_{b\neq a}^{\ntrksignal}     \ddelta +
    \sum_{a}^{\ntrkbackground}\sum_{b\neq a}^{\ntrkbackground} \ddelta \nonumber \\
&+& \sum_{a}^{\ntrkbackground}\sum_{b}^{\ntrksignal}           \ddelta + 
    \sum_{a}^{\ntrksignal}    \sum_{b}^{\ntrkbackground}       \ddelta, 
\end{eqnarray}
where the indices $a$ and $b$ run over tracks in a subevent of its 
  corresponding category, $\Delta\phi^{ab}$ is a short-hand notation for
  $\phi^a-\phi^b$, and the Dirac delta function \ddelta\ ensures that the requirement 
  $\dphi=\phi^a-\phi^b$ is satisfied.
Besides requiring that the index $b \neq a$, as is made explicit in 
  Eq.~\eqref{eq:corr_long} above, the requirement that 
  $|\eta^a-\eta^b|>2$ is also imposed.
This requirement can be imposed in Eq.~\eqref{eq:corr_long} by including
  the step function $\Theta(|\eta^a-\eta^b|-2)$, but for brevity, 
  is not included explicitly.
Additionally the indices $a$ and $b$ are restricted to the particles 
  within the chosen \pt-ranges for the reference and associated 
  particles, respectively.

To take account of different pile-up conditions, the analysis is done in intervals of \avgntrk. 
Therefore, the expression given by Eq.~\eqref{eq:corr_long} has to be 
  summed over a subset of data in each \avgntrk interval. 
In the following, the number of events in the interval where the number of observed tracks 
  is \ntrkdirect, is denoted by $\nevtdir$. 
Averaging the first contribution in Eq.~\eqref{eq:corr_long} over all events 
  at fixed \ntrkdirect\ and \avgntrk yields:
\begin{eqnarray}
\frac{1}{\nevtdir}
\sum_{n}^{\nevtdir}\sum_{a}^{\ntrksignal}\sum_{b\neq a}^{\ntrksignal} \ddelta
&=&
\sum_{\ntrksignal=0}^{\ntrkdirect}\probsig\sal\frac{\dnpairsig}{\ddphisig}\big(\ntrksignal\big)\sar \equiv
\dal\frac{\dnpairsig}{\ddphisig}\big(\ntrksignal\big)\dar.
\label{eq:corr_term1}
\end{eqnarray}

In the presence of pile-up, the contributions to the \Direct tracks
  come from different numbers of \Signal tracks such 
	that $\ntrksignal\leq\ntrkdirect$ (as $\ntrksignal+\ntrkbackground=\ntrkdirect$). 
Probabilities to find \ntrksignal in events are denoted \probsig\ and 
  are shown in Figure~\ref{fig:refold}.
For clarity, the parameters that this probability depends on, 
  i.e. \avgntrk and \ntrkdirect, are labelled explicitly here.
The averaging is done over all values of \ntrksignal, which is 
  reflected by the double angular bracket that appears in the equation: 
  the average over events with fixed \ntrksignal is denoted by the smaller angular brackets, and the weighted 
  average over all \ntrksignal for a given \ntrkdirect in a category is denoted by larger angular brackets. 
In practice, only a relatively narrow region of \ntrksignal effectively contributes to 
  $\dnpairsig/\ddphisig$.  
The width of this region depends on \ntrkdirect and on \avgntrk.

Similarly to the first contribution, the second contribution to Eq.~\eqref{eq:corr_long} can be written as:
\begin{eqnarray}
\frac{1}{\nevtdir}
\sum_{n}^{\nevtdir}
\sum_{a}^{\ntrkbackground}
\sum_{b\neq a}^{\ntrkbackground}
\ddelta =
\sum_{\ntrksignal=0}^{\ntrkdirect}\probsig\
\sal\frac{\dnpairbkg}{\ddphibkg}\big(\ntrkbackground\big)\sar
=\dal\frac{\dnpairbkg}{\ddphibkg}\big(\ntrkbackground\big)\dar.
\label{eq:corr_term2}
\end{eqnarray}

Averaging the last two terms in Eq.~\eqref{eq:corr_long} over the 
  event sample eliminates any $\dphi$ dependence except a 
  constant one, because the \Bkg tracks cannot be correlated with 
  \Signal tracks since they originate from different interactions. 
The third term in Eq.~\eqref{eq:corr_long} can be written as:
\begin{eqnarray}
\frac{1}{\nevtdir}\sum_{n}^{\nevtdir} \bigg(\sum_{a}^{\ntrkbackground}\sum_{b}^{\ntrksignal    } \ddelta\bigg) = \nonumber \\
 \sum_{\ntrksignal=0}^{\ntrkdirect}\probsig\iint\sal \frac{\dnsingbkg}{\dphibkga}\big(\ntrkbackground\big) \sar \sal \frac{\dnsingsig}{\dphisigb}\big(\ntrksignal\big) \sar\ddelta\mathrm{d}\phi^a\mathrm{d}\phi^b = \nonumber \\
 \dals{\iint} \frac{\dnsingbkg}{\dphibkga}\big(\ntrkbackground\big) \sar \sal \frac{\dnsingsig}{\dphisigb}\big(\ntrksignal    \big) \dars{\ddelta\mathrm{d}\phi^a\mathrm{d}\phi^b},
\label{eq:corr_term3}
\end{eqnarray}
where $\big\langle\dnsing/\dphi\big\rangle$ are the single-particle 
  angular track densities averaged over many events.
Equation~\eqref{eq:corr_term3} states that averaged over many events, 
  the pair distribution involving \Signal\ and \Bkg\ tracks
  can be replaced by the convolution of the individual 
  single-particle distributions.
The fourth term in Eq.~\eqref{eq:corr_long} gives an expression 
  identical to Eq.~\eqref{eq:corr_term3} except that the indices $a$
  and $b$ interchanged.
Substituting Eqs.~\eqref{eq:corr_term1}--\eqref{eq:corr_term3} into 
  Eq.~\eqref{eq:corr_long} and rearranging gives:
\begin{eqnarray}
\dal\frac{\dnpairsig}{\ddphisig}\big(\ntrksignal\big) \dar
&=&
\sal\frac{\dnpairdir}{\ddphidir}\big(\ntrkdirect\big) \sar -\ 
\dal\frac{\dnpairbkg}{\ddphibkg}\big(\ntrkbackground\big)\dar  \nonumber \\
&-&
\dals{\iint}\frac{\dnsingbkg}{\dphibkga}\big(\ntrkbackground\big)\sar \sal\frac{\dnsingsig}{\dphisigb}\big(\ntrksignal\big)\dars{\ddelta\mathrm{d}\phi^a\mathrm{d}\phi^b} \nonumber \\ 
&-& 
\dals{\iint}\frac{\dnsingsig}{\dphisiga}\big(\ntrksignal\big)\sar \sal\frac{\dnsingbkg}{\dphibkgb}\big(\ntrkbackground\big)\dars{\ddelta\mathrm{d}\phi^a\mathrm{d}\phi^b}.
\label{eq:corr_exact}
\end{eqnarray}

So far, no approximations are made in the derivation of 
  Eq.~\eqref{eq:corr_exact}. 
However, in implementing the pile-up subtraction in the analysis, 
  some approximations are necessary. 

The first approximation relies on $\Bkg\equiv\Mixed$, which is 
  established earlier in the analysis.
Then the \Bkg\ terms in Eq.~\eqref{eq:corr_exact} can
  be replaced by the corresponding \Mixed\ terms.
Additionally the single-track distributions for \Signal tracks on the
  second line of Eq.~\eqref{eq:corr_exact} can be written as:
\begin{eqnarray*}
\sal\frac{\dnsingsig}{\dphisiga}\big(\ntrksignal    \big)\sar=
\sal\frac{\dnsingdir}{\dphidira}\big(\ntrkdirect    \big)\sar-
\sal\frac{\dnsingmix}{\dphimixa}\big(\ntrkbackground\big)\sar.
\end{eqnarray*}
With these substitutions, Eq.~\eqref{eq:corr_exact} becomes:
\begin{eqnarray}
\dal\frac{\dnpairsig}{\ddphisig}\big(\ntrksignal\big) \dar\ 
&=&\
\sal\frac{\dnpairdir}{\ddphidir}\big(\ntrkdirect\big) \sar -\ 
\dal\frac{\dnpairmix}{\ddphimix}\big(\ntrkbackground\big) \dar  \nonumber \\
&-& \dals{\iint}\frac{\dnsingmix}{\dphimixa}\big(\ntrkbackground\big)\sar \sal \frac{\dnsingdir}{\dphidirb}\big(\ntrkdirect    \big) \dars{\ddelta\mathrm{d}\phi^a\mathrm{d}\phi^b} \nonumber\\ 
&-& \dals{\iint}\frac{\dnsingdir}{\dphidira}\big(\ntrkdirect    \big)\sar \sal \frac{\dnsingmix}{\dphimixb}\big(\ntrkbackground\big) \dars{\ddelta\mathrm{d}\phi^a\mathrm{d}\phi^b} \nonumber\\
&+& \dals{\iint}\frac{\dnsingmix}{\dphimixa}\big(\ntrkbackground\big)\sar \sal \frac{\dnsingmix}{\dphimixb}\big(\ntrkbackground\big) \dars{\ddelta\mathrm{d}\phi^a\mathrm{d}\phi^b} \nonumber\\
&+& \dals{\iint}\frac{\dnsingmix}{\dphimixa}\big(\ntrkbackground\big)\sar \sal \frac{\dnsingmix}{\dphimixb}\big(\ntrkbackground\big) \dars{\ddelta\mathrm{d}\phi^a\mathrm{d}\phi^b}.
\label{eq:corr_approx1}
\end{eqnarray}

The second approximation requires that $\dnpairsig/\ddphisig$ 
  changes slowly with \ntrksignal, i.e. that the correlations do not 
  change significantly over the range of \ntrksignal that contributes 
  to a given \ntrkdirect. 
In other words, this assumption requires that the analysed correlation 
  does not change significantly over an effective range of \ntrksignal 
  that cannot be resolved in the presence of the pile-up. 
Those ranges are effectively the widths of the peaks shown in 
  Figure~\ref{fig:refold}, and are fixed for a given background 
  condition \avgntrk.
By limiting the background condition to $\avgntrk < \avgntrk_{\mathrm{max}}$, 
  one can control the magnitude of this width. 
In the present analysis, the maximum value of the background condition
  is chosen to be $\avgntrk_{\mathrm{max}}=7.5$. This limit is shown in panel (c) of Figure~\ref{fig:mains}.

To measure the two-particle correlation as function of \ntrksignal in 
  the presence of pile-up, quantities defined by 
  Eq.~\eqref{eq:corr_approx1} found at fixed values of \ntrkdirect and 
  in different intervals of \avgntrk have to be summed with 
  weights as: 
\begin{eqnarray}
\frac{\dnpairsig}{\ddphisig}\big(\ntrksignal\big) \approx 
\frac{1}{\sum_{\avgntrk<\avgntrk_{\mathrm{max}}} \nevtdir}\sum_{\avgntrk<\avgntrk_{\mathrm{max}}}\sum_{\ntrkdirect\geq\ntrksignal} \nevtdir\probsig\dal\frac{\dnpairsig}{\ddphisig}\big(\ntrksignal\big) \dar .
\label{eq:corr_def}
\end{eqnarray}

Combining Eqs.~\eqref{eq:corr_approx1}~and~\eqref{eq:corr_def} the final 
  result is obtained using the expression: 
\begin{eqnarray}
\frac{\dnpairsig}{\ddphisig}\big(\ntrksignal\big) 
&\approx&
\frac{1}{\sum_{\avgntrk}\nevtdir}\sum_{\avgntrk}\sum_{\ntrkdirect\geq\ntrksignal} \nevtdir\probsig
\bigg(
\sal\frac{\dnpairdir}{\ddphidir}\big(\ntrkdirect\big) \sar 
-
\bigg[
\dal\frac{\dnpairmix}{\ddphimix}\big(\ntrkbackground\big) \dar \nonumber \\
&+& \dals{\iint}\frac{\dnsingmix}{\dphimixa}\big(\ntrkbackground\big)\sar \sal \frac{\dnsingdir}{\dphidirb}\big(\ntrkdirect    \big) \dars{\ddelta\mathrm{d}\phi^a\mathrm{d}\phi^b} \nonumber \\
&+& \dals{\iint}\frac{\dnsingdir}{\dphidira}\big(\ntrkdirect    \big)\sar \sal \frac{\dnsingmix}{\dphimixb}\big(\ntrkbackground\big) \dars{\ddelta\mathrm{d}\phi^a\mathrm{d}\phi^b} \nonumber \\
&-& \dals{\iint}\frac{\dnsingmix}{\dphimixa}\big(\ntrkbackground\big)\sar \sal \frac{\dnsingmix}{\dphimixb}\big(\ntrkbackground\big) \dars{\ddelta\mathrm{d}\phi^a\mathrm{d}\phi^b} \nonumber \\
&-& \dals{\iint}\frac{\dnsingmix}{\dphimixa}\big(\ntrkbackground\big)\sar \sal \frac{\dnsingmix}{\dphimixb}\big(\ntrkbackground\big) \dars{\ddelta\mathrm{d}\phi^a\mathrm{d}\phi^b}
\bigg]
\bigg) .
\label{eq:corr_ana}
\end{eqnarray}

The analysis uses Eq.~ \eqref{eq:corr_ana} in the following way.
In each category of \avgntrk, the distributions of two-particle 
  pair-distributions are built for all values of \ntrkdirect and for 
  all values of \ntrkmixed. 
They are then summed using weights \probsig\ to build the 
  background contributions, given by the square brackets in 
  Eq.~\eqref{eq:corr_ana} for different \ntrkdirect and 
  $\ntrkmixed =\ntrkdirect-\ntrksignal$ combinations. 
Next, these contributions are subtracted from the distribution measured  
 in the \Direct event for each \ntrkdirect, giving the expression 
  in round brackets. 
Subtracted results are weighted with probabilities \probsig\ and 
  multiplied by $\nevtdir$, the number of events 
  with any given \ntrkdirect. The resulting distributions are added to 
  the distributions of \Signal events for all values such 
  that $\ntrksignal\leq\ntrkdirect$. 
In the last step, the values of 
  $\dnpairsig/\ddphisig$ in those 
  categories of \avgntrk that are used in the analysis are added 
  together and normalised.

Equation~\eqref{eq:corr_ana} gives the pile-up-corrected
  distribution of track-pairs --- $S(\dphi)$ in Eq.~\eqref{eq:ana0} ---
  evaluated at fixed \ntrksignal.
The pair-acceptance distribution $B(\dphi)$ does not require any 
  correction for pile-up as it is an estimate of the detector acceptance
  which is not affected by pile-up.
The pile-up-corrected correlation functions $C(\dphi)$ are then built 
  by dividing the $S(\dphi)$ by the $B(\dphi)$ and normalizing to a 
  $\dphi$-averaged value of unity.
\section{Template fits}
\label{sec::template_fits}
Figure~\ref{fig:template_fits_all} shows the pile-up-corrected 2PC for 
  several \ntrks intervals for the 13~\TeV\ \Zboson-tagged data. 
Correlations are measured for tracks in the $0.5<\ptab<5$\,\GeV\ 
  range. 
In the higher track multiplicity intervals, a clear enhancement on the
  near-side ($\dphi=0$) is visible.
Figure~\ref{fig:template_fits_all} also shows results for the template 
fits (Eq.~\eqref{eq:template}) to the 2PC, with the \ntrks interval of $20<\ntrks\leq30$ 
  used as the peripheral reference.
The measured correlation functions are well described 
  by the template fits, and long-range correlations (indicated by 
  dashed blue lines) are observed.
The fits in Figure~\ref{fig:template_fits_all} include harmonics $n=2$--4,
  however the subsequent analysis described in this paper
  focusses only on $\vtwo$, as the associated systematic and statistical 
  uncertainties on the higher order harmonics are quite large. 
\begin{figure}[t]
\begin{centering}
\includegraphics[width=1\linewidth]{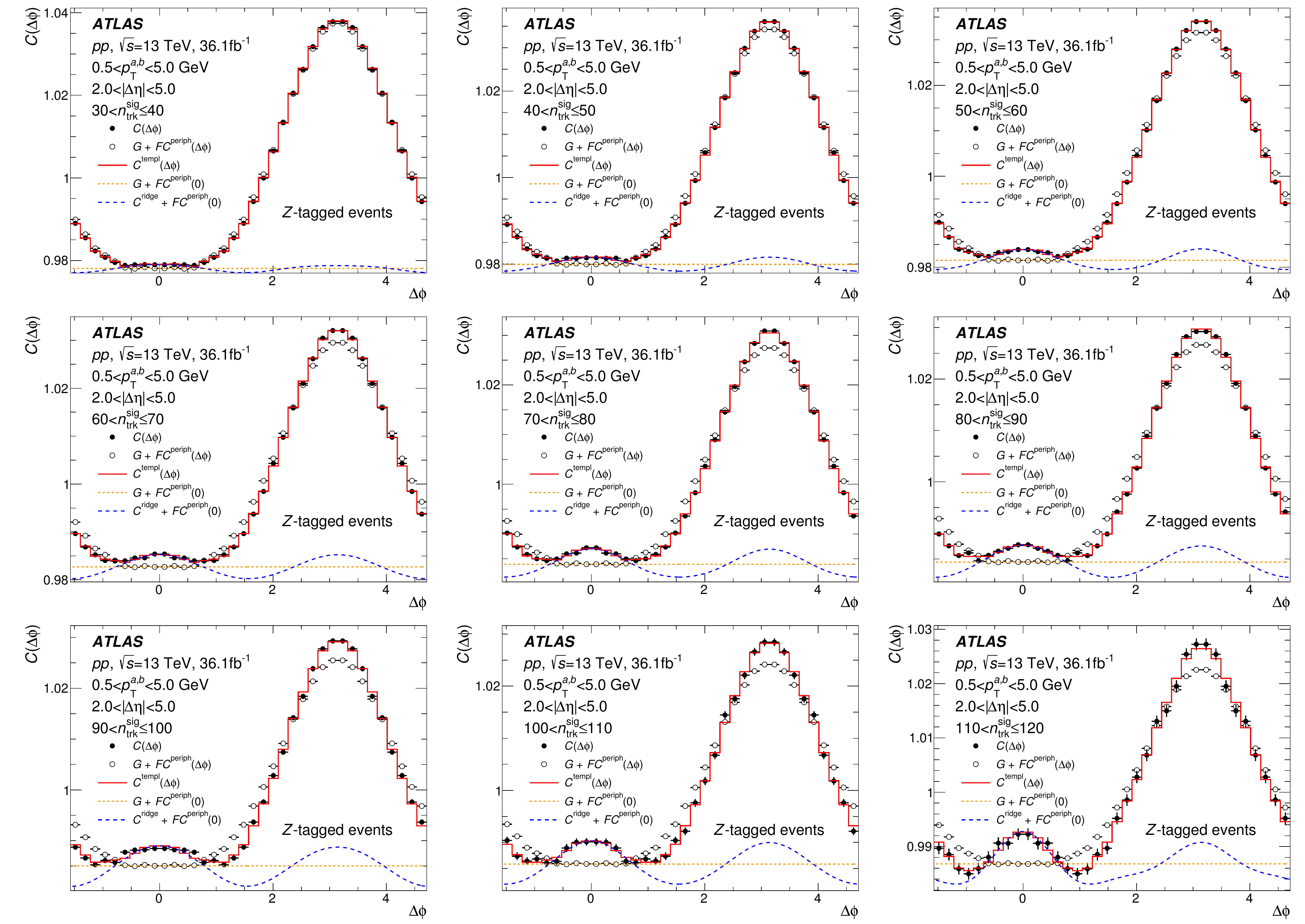}
\end{centering}
\caption{
Template fits to the pile-up-corrected $C(\dphi)$ in the 
  13~\TeV\ \Zboson-tagged data.
The different panels correspond to different 
  \ntrks\ intervals. 
The $\ntrks\in(20,30]$ interval is used to determine the $\cperi(\dphi)$,
  and the template fits include harmonics $n=2$--4.
The $F\cperi(\dphi)$ and $C^{\mathrm{ridge}}$ terms have been shifted
  up by $G$ and $F\cperi(0)$ respectively, for easier comparison.
The plots are for $0.5<\ptab<5$\,\GeV.
\label{fig:template_fits_all}}
\end{figure}

From the template fits the \vtwo is extracted following 
  Eq.\,\eqref{eq:factorize}.
The left panel of Figure~\ref{fig:template_fits_v2} shows the \vtwo values 
  obtained from the template fits as a function of \ntrks.
The \vtwo values before correcting for pile-up are also shown 
  for comparison. 
  Without the pile-up correction, a clear monotonic decrease in \vtwo is observed with increasing track multiplicity, corresponding to an increase in pile-up contamination.

The right panels of Figure~\ref{fig:template_fits_v2} show the ratios 
  of uncorrected \vtwo\ to the 
  corresponding values with the pile-up correction.
The uncorrected values show a significant decrease with increasing 
  multiplicity and are $\sim$25\% (20\%) lower than the corrected one in 
  8~\TeV\ (13~\TeV) data at the highest measured track multiplicity.
However, after the pile-up correction, the \vtwo shows a significantly weaker
  dependence on the track multiplicity, similar to the 
  observations in Refs.\,\cite{HION-2015-09,HION-2016-01}).

\begin{figure}[h]
\begin{centering}
\includegraphics[width=1\linewidth]{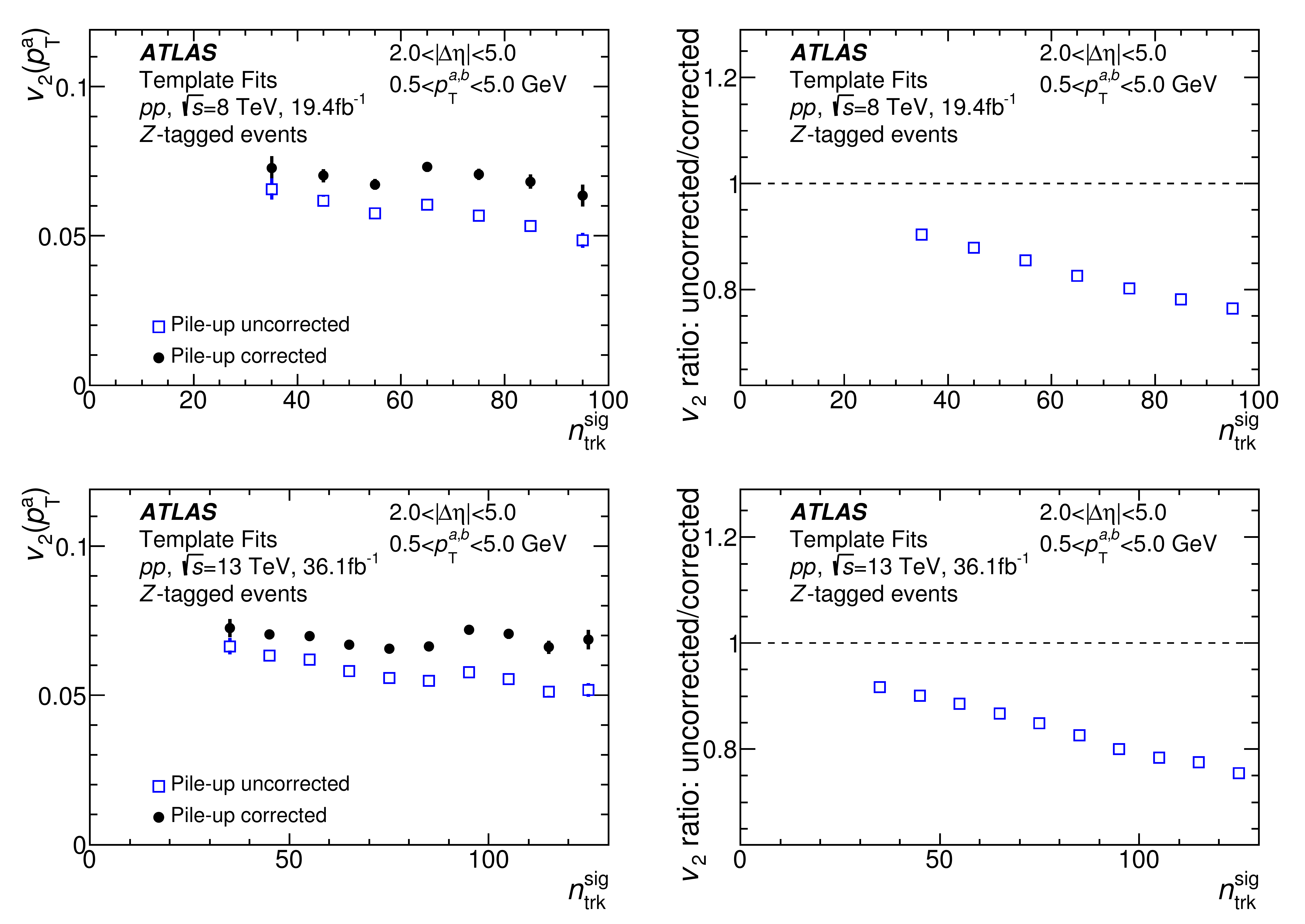}
\end{centering}
\caption{
Top left panel shows the \vtwo values obtained from the template fits 
  in the 8~\TeV\ data, corrected for pile-up, plotted as a 
  function of the \ntrksignal (black points).
For comparison, the \vtwo not corrected for pile-up is also plotted. 
The uncorrected \vtwo\ is also plotted as a function of \ntrksignal -- 
  the pile-up corrected multiplicity -- so that the effect of the
  pile-up correction on the \vn\ is compared between the same set of events.
The top right panel shows the ratio of the two \vtwo values.
Bottom row shows similar plots for the 13~\TeV\ data.
The error bars indicate statistical uncertainties and are not shown in the 
  ratio plots.
Plots are for $0.5<\ptab<5$~\GeV.
\label{fig:template_fits_v2}}
\end{figure}

Figure~\ref{fig:v2_aux_pile-up_correction_pt} compares the \pt\ 
  dependence of the \vtwo\ before and after correcting for pile-up.
The \pt dependence is evaluated over a broad 40--100 \ntrks\ range.
A dependence of the correction on the \pt is observed.
Over the 0.5--3~\GeV\ \pT\ interval, the magnitude of the correction 
  decreases with increasing \pT.

\begin{figure}[!h]
\centering
\includegraphics[width=1\linewidth]{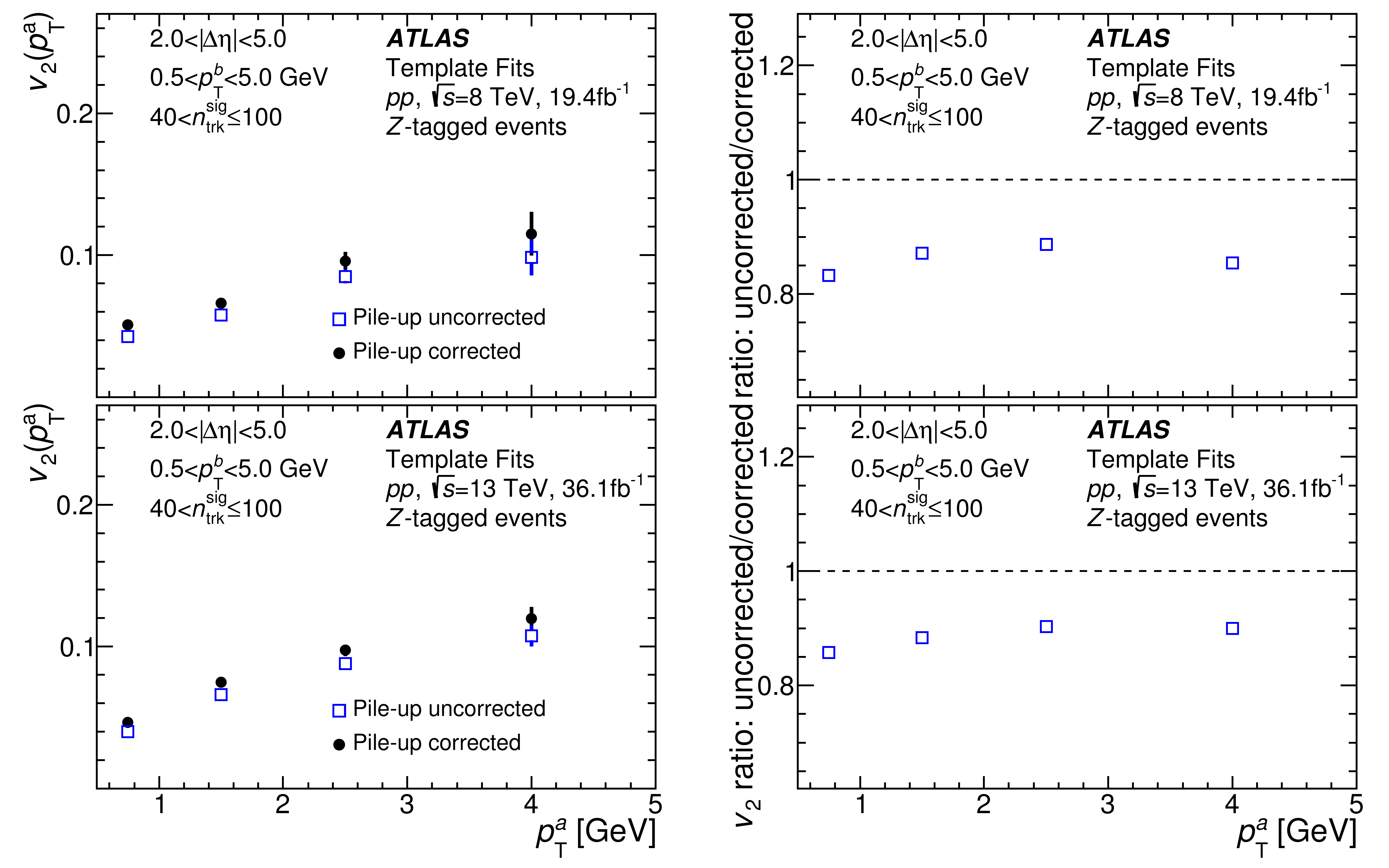}
\caption{
Top left panel compares the \vtwo\ values as a function of \pt
  in the 8~\TeV\ data, when correcting (solid-black points) and 
  not correcting (open blue points) for pile-up pairs.
The plots are for the 40--100 \ntrksignal interval.
The blue points correspond to the Fourier coefficients of the 
  uncorrected 2PCs for $40<\ntrksignal\leq100$.
The top right panel shows the ratio of the uncorrected \vtwo to the 
  corrected \vtwo.
The error bars indicate statistical uncertainties and are not shown in the 
  ratio plots.
The lower panels show similar plots for the 13~\TeV\ data.
\label{fig:v2_aux_pile-up_correction_pt}
} 
\end{figure}

\section{Systematic uncertainties}
\label{sec:syst}

The systematic uncertainties in the \vtwo\ measurement can broadly be classified into two categories: the first
  category comprises systematic uncertainties that are intrinsic to the 2PC and to the template-fitting
   procedure and have been used in previous 2PC analyses~\cite{HION-2015-09,HION-2016-01}.
These include uncertainties from the choice of peripheral bin used in the
  template fits, the tracking efficiency, and the pair-acceptance.
The second category comprises the uncertainties associated with the correction
  of the \vtwo that accounts for pile-up tracks; these uncertainties are specific to the
  present analysis.

\subsection{Peripheral interval}
The template-fitting procedure~\cite{HION-2015-09,HION-2016-01} uses the $\ntrks\in$(20,30] interval as
  the peripheral reference.
To test the sensitivity of the measured \vtwo to any residual changes
  in the width of the away-side ($\dphi=\pi$) jet peak and to the \vtwo present in
  the peripheral reference, the analysis is repeated using the 0--20,
  10--20, and 30--40 multiplicity intervals as the peripheral reference.
The resulting variation in the \vtwo\ when using these alternative
  peripheral references is included as a systematic uncertainty.
The assigned uncertainties are conservatively taken to be 
  larger of the three variations and symmetric about the nominal value.
For the multiplicity dependence of the \vtwo\ measured in the
  integrated \pt\ interval of 0.5--5~\GeV\, this uncertainty varies
  from $\sim$8\% at $\ntrks=30$ to $\sim$3\% for $\ntrks>70$
  in the 8~\TeV\ data.
For the 13~\TeV\ data the uncertainty is within 4\% across the entire
  measured multiplicity range.
For the \pt dependence, this uncertainty varies from 4\% to 15\%
  depending on the \pt and the dataset.

\subsection{Track reconstruction efficiency}
In evaluating the correlation functions, each particle is weighted by
 a factor $1/\epsilon(\pt,\eta)$ to account for the tracking efficiency.
The systematic uncertainties in the efficiency $\epsilon(\pt,\eta)$
  thus need to be propagated into \ctwophi and the final
  \vtt measurements.
The \ctwophi and \vtwo are mostly insensitive to the tracking efficiency.
This is because the \vtt measures the relative variation of the yields
  in \dphi; an overall increase or decrease in the efficiency changes
  the yields but does not affect the \vtwo.
However, due to \pT and $\eta$ dependence of the tracking efficiency
  and its uncertainties~\cite{STDM-2015-02}, there is some residual effect on the \vtwo.
The corresponding uncertainty in the \vtwo is  estimated by repeating
  the analysis while varying the efficiency within its upper and lower
  uncertainty values --- of about 5\% --- in a \pT-dependent manner.
For \vtwo this uncertainty is estimated to be less than 1\%, when studying
  the multiplicity dependence for the 0.5--5~\GeV\ \pt\ interval, and
  less than 0.5\% for the differential $\vtwo(\pt)$.

\subsection{Pair-acceptance}
The analysis relies on the $B(\dphi)$ distribution to correct for the
  pair-acceptance  of the detector using Eq.\,\eqref{eq:ana0}.
The $B(\dphi)$ distributions are nearly flat in $\dphi$, and the effect
  on the \vtwo when correcting for the acceptance is less than 1\%
  for all multiplicities and \pt.
Since the pair-acceptance corrections are small, the entire correction
  is conservatively taken as the systematic uncertainty associated with
  pair-acceptance effects.

\subsection{Accuracy of the background estimator}
This uncertainty arises due to inaccuracy in the determination of \nint during the run and stability of the \zvtx distribution. They are estimated using the inaccuracy in the luminosity determination described in Ref.~\cite{Aaboud:2016hhf,ATLAS-CONF-2019-021} and stability studies performed in the analysis. Another contribution is coming from the quality of the fits. Although fits used in the functional form of Eq.\,\eqref{eq:bkg} accurately reproduce data as shown in Figures~\ref{fig:pv_pointing}~and~\ref{fig:fit_results},  alternative fit functions are also studied to derive an uncertainty that, together with the factors mentioned earlier, results in $\lesssim$1\% uncertainty added to the final results.

\subsection{Uncertainties in transition matrices}
The transition matrices discussed in Section\,\ref{sec:unfold} for unfolding
  the \ntrk distributions and for finding coefficients for correcting the 2PC are determined
  using data.
The \ntrks distribution is approximated with the \Direct distributions in the
  lowest $\avgntrk$ interval ($\avgntrk<0.5$).
Uncertainties in the \vtwo values due to this approximation are estimated by repeating the analysis
  with the matrices calculated from the \Direct distributions in the interval $\avgntrk<1$.
The variation is less than 2\% throughout, and is included as a
 systematic uncertainty in the value of \vtwo.

\subsection{Accuracy of the pile-up correction procedure}
As described in Section\,\ref{sec:pile-up}, the pile-up correction
 procedure is implemented in intervals of \avgntrk.
In order to check residual pile-up effects that are not removed by the correction
  procedure, a study of the pile-up-corrected \vtwo is performed as a
  function of \avgntrk, and the variation in the measured \vtwo\ is included
  as a systematic uncertainty.
This uncertainty is determined to be $\pm3.5$\% for the 8~\TeV\ data
  across the measured multiplicity range.
For the 13~\TeV\ data, this uncertainty is $\pm4$\% for $\ntrksignal<100$
  but increases to 15\% at higher multiplicities.

An independent check of the pile-up correction procedure is done by performing
  an MC closure analysis using the MC sample described in Section~\ref{sec:mixed}.
Since the MC generated events do not have any physical long-range correlations,
  the closure test is performed on the Fourier components of the 2PC defined
  by Eq.~\eqref{eq:fourier_fit}--\eqref{eq:factorize} instead.
The Fourier-\vtwo includes mostly contributions from back-to-back dijets,
  which is the main (physics) background that the template analysis removes.
The pile-up affects the 2PC caused by the dijet and the long-range correlations in a similar way.
Thus the Fourier-\vtwo is an ideal object to use in checking the performance of the pile-up correction, because it has a
non-zero value in the simulated events. The MC closure test is performed as follows. The generator-level \vtwo
 is obtained from 2PCs using only reconstructed tracks that are known to be associated with the \Zboson-boson
 vertex. The generated \vtwo is then compared to the reconstruction-level Fourier-\vtwo obtained when applying
 the pile-up correction procedure used in the analysis of the real data. The corrected and generated \vtwo are
 found to be consistent within the systematic uncertainties associated with the pile-up correction procedure.

Table\,\ref{tab:List-of-systematics} summarizes the systematic
  uncertainties in the multiplicity dependence of the measured \vtwo.
Table~\ref{tab:List-of-systematics-pt} summarizes the uncertainties
  for the \pt dependence of \vtwo.
The dominant systematic uncertainties arise from the choice of peripheral reference
  and from the \avgntrk\ dependence.
\begin{table}[htb!]
\begin{centering}
\caption{
Systematic uncertainties for the multiplicity dependence of the \vtwo
  integrated over the 0.5--5~\GeV\ \pt interval.
Where ranges are provided for both multiplicity and the uncertainty,
  the uncertainty varies from the first value to the second value
  as the multiplicity varies from the lower to upper limits of the range.
The listed uncertainties are taken to be symmetric about the nominal \vtwo values.
\label{tab:List-of-systematics}}
\begin{tabular}{l|cS[table-format=4.1]|cS[table-format=4.1]}
   & \multicolumn{2}{c|}{8~\TeV} & \multicolumn{2}{c}{13~\TeV}              \tabularnewline
Source           & \ntrksignal  &\multicolumn{1}{c|}{Uncertainty [\%]}   &
          \ntrksignal  &\multicolumn{1}{c}{Uncertainty [\%]} \tabularnewline \hline 
Choice of peripheral bin                 & 30--70       & \multicolumn{1}{c|}{8--3}   &30--90   &4.0  \tabularnewline
                                         & 70--100      & 3          &90--140    &3.5            \tabularnewline
Tracking efficiency                      & 30--100      & 1          &30--140    &1    \tabularnewline
Pair acceptance                          & 30--100      & 1          &30--140    &0.5  \tabularnewline
Accuracy of \avgntrk estimation          & 30--100      & 1          &30--140    &0.5  \tabularnewline
Uncertainties in transition matrices     & 30--100      & 2          &30--140    &1.5  \tabularnewline
\avgntrk dependence                      & 30--100      & 3.5        &30--100    &4    \tabularnewline
                                         &              &            &100--120   &6    \tabularnewline
                                         &              &            &120--140   &15   \tabularnewline
\end{tabular}
\par\end{centering}
\end{table}
\begin{table}[h]
\begin{centering}
\caption{
Systematic uncertainties for the $\vtwo(\pt)$.
\label{tab:List-of-systematics-pt}}
\begin{tabular}{l|cS[table-format=4.1]|cS[table-format=4.1]}
                                        & \multicolumn{2}{c|}{8~\TeV}              & \multicolumn{2}{c}{13~\TeV}    \tabularnewline
Source                                  & \pT~[\GeV]         &\multicolumn{1}{c|}{Uncertainty [\%]}
                                        & \pT~[\GeV]         &\multicolumn{1}{c}{Uncertainty [\%]} \tabularnewline \hline
Choice of peripheral bin                & 0.5--3             & 4                   &0.5--1   & 7   \tabularnewline
                                        & 3--5               & 7                   &1--2     & 5   \tabularnewline
                                        &                    &                     &2--5     &15   \tabularnewline
Tracking efficiency                     & 0.5--5             & 0.5                 &0.5--5   &0.5  \tabularnewline
Pair acceptance                         & 0.5--5             & 1                   &0.5--5   & 1   \tabularnewline
Accuracy of \avgntrk estimation         & 0.5--5             & 0.5                 &0.5--5   &0.5  \tabularnewline
Uncertainties in transition matrices    & 0.5--5             & 1                   &0.5--5   &0.5  \tabularnewline
\avgntrk dependence                     & 0.5--5             & 3.5                 &0.5--5   &4    \tabularnewline
\end{tabular}
\par\end{centering}
\end{table}

\section{Results \label{sec:results}}
Figure\,\ref{fig:template_fits_v2_final} shows the final results for the
  multiplicity dependence of the \vtwo with all systematic
  uncertainties included.
The results are corrected to account for pile-up and detector efficiency effects,
  and are plotted as a function of the measured multiplicity (\ntrks),
  which is corrected for pile-up, but not
  for detector efficiency effects.\footnote{
    Integrated over $\pt>0.4$~\GeV, the tracking efficiency is on
      average $\sim$6\% lower (absolute) for the 8~\TeV\ \Zboson-tagged data
      compared to the other data shown in Figure~\ref{fig:template_fits_v2_final}.
    Therefore, the same \ntrks\ corresponds to slightly higher true multiplicity for
      the 8~\TeV\ data.
  }
The left panel compares the final \vtwo values obtained from
  the template fit in the 8 and 13~\TeV\ \Zboson-tagged samples, to the
  \vtwo values obtained in 5 and 13~\TeV\ inclusive
  \pp\ collisions from Ref.\,\cite{HION-2016-01}.
The right panel shows the ratio of the \vtwo\ in the 8 and 13~\TeV\
  \Zboson-tagged samples to the \vtwo\ in 13~\TeV\ inclusive
  \pp\ collisions.
The systematic uncertainties in the measured \vtwo\ for a given
  dataset are to some extent correlated across the different multiplicity intervals 
  shown in Figure\,\ref{fig:template_fits_v2_final}.
The \Zboson-tagged \vtwo values show no significant dependence on the
  multiplicity, similar to the results obtained from the inclusive
  samples, and are consistent with each other as well as the inclusive
  measurements, within 1--2$\sigma$ systematic uncertainties.

Figure~\ref{fig:template_fits_v2_final_pt} compares the \pt dependence
  of the template-\vtwo\ between the \Zboson-tagged and inclusive
  measurements.
The \pt\ dependence is evaluated over the 40--100 \ntrks\ range.
The upper limit of $100$ in the chosen multiplicity range is because the 
  8~\TeV\ \Zboson-tagged measurements are done upto this multiplicity.
The lower limit of 40 tracks arises because the 30--40 track multiplicity 
  range is used as the peripheral reference in the systematic uncertainty 
  estimates, and thus events with less than 40 tracks are excluded 
  from the measurement.
As seen for the multiplicity dependence, the \pt dependence is
  also consistent between the 8 and 13~\TeV\ \Zboson-tagged samples.
The \Zboson-tagged $\vtwo(\pt)$ values are also consistent with the
  inclusive measurements within 1--1.5$\sigma$ systematic
  uncertainties.

\begin{figure}[h]
\begin{centering}
\includegraphics[width=1\linewidth]{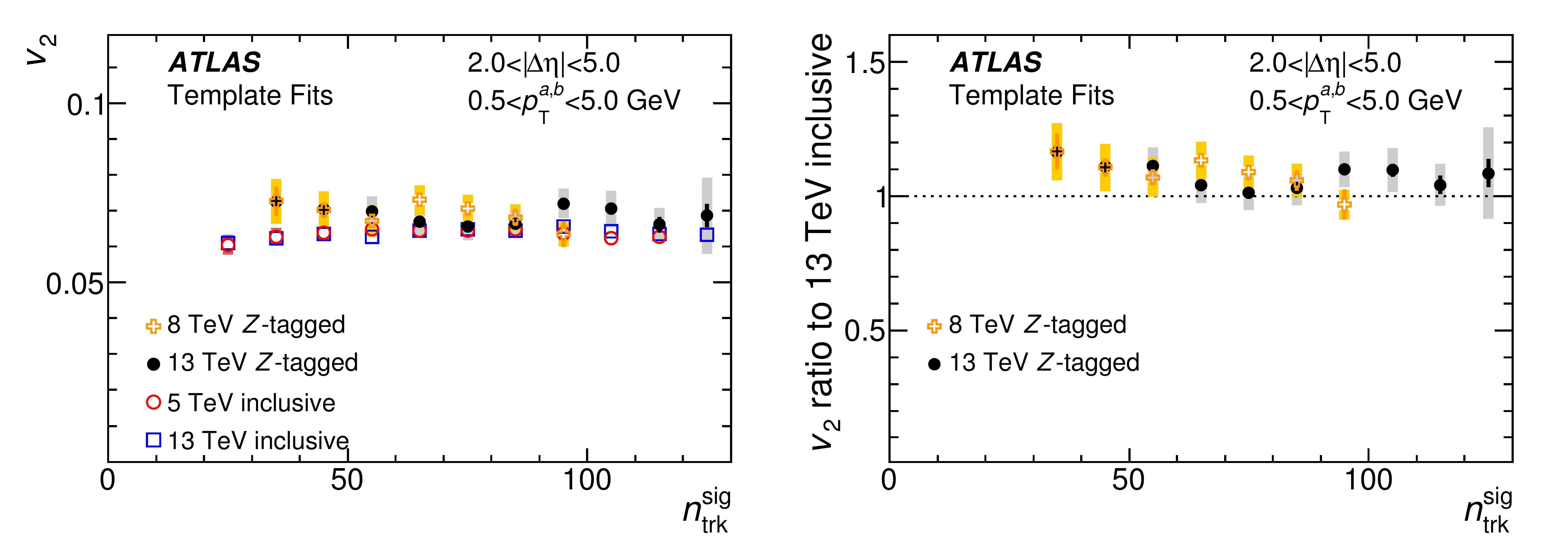}
\end{centering}
\caption{
Left panel: the pile-up-corrected \vtwo\ values obtained from the
  template fits as a function of \ntrks.
For comparison, the \vtwo\ values obtained in 5 and 13~\TeV\ inclusive
  \pp\ data from Ref.\,\cite{HION-2016-01} are also shown.
The error bars and shaded bands indicate statistical and systematic
  uncertainties, respectively.
Right panel: the ratio of the \vtwo in 8 and 13~\TeV\ \Zboson-tagged
  samples to the \vtwo\ in inclusive 13~\TeV\ \pp\ collisions as a
  function of \ntrks.
The horizontal dotted line indicates unity and is intended to guide the eye.
Results are ploted for $0.5<\ptab<5$~\GeV.
\label{fig:template_fits_v2_final}}
\end{figure}

\begin{figure}[h]
\begin{centering}
\includegraphics[width=1\linewidth]{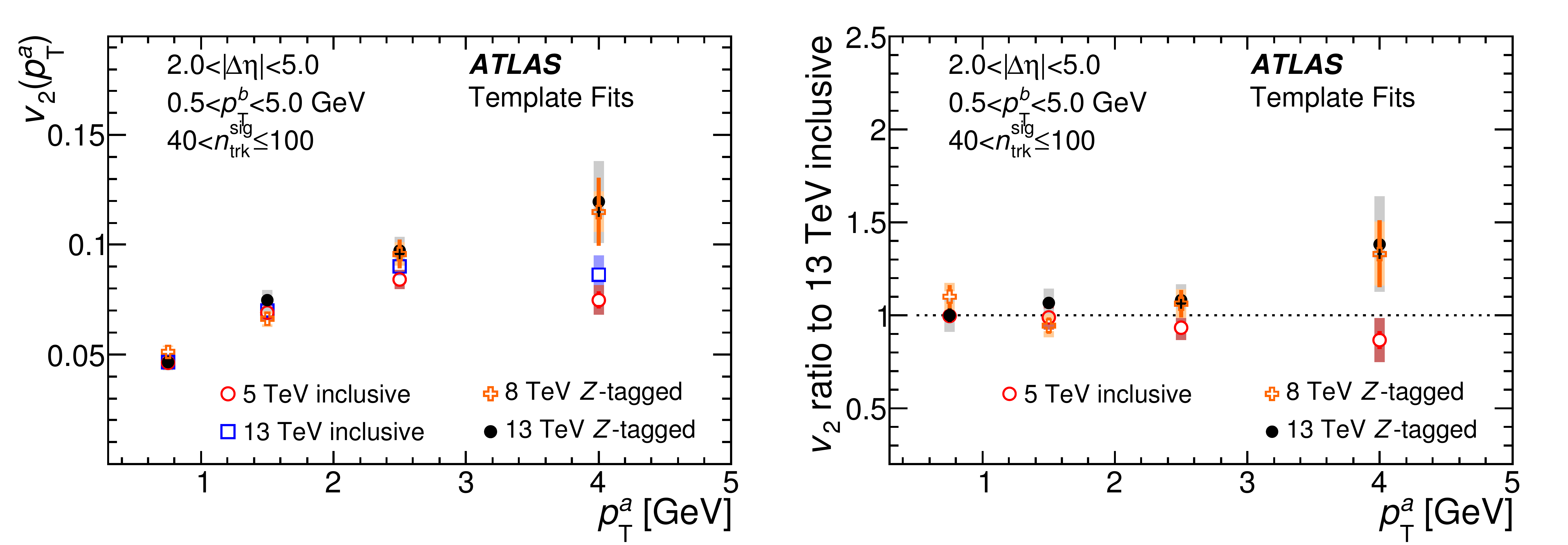}
\end{centering}
\caption{
Left panel: the pile-up-corrected \vtwo\ values obtained from the
  template fits as a function of \pt.
For comparison, the \vtwo\ values obtained in 5 and 13~\TeV\ inclusive
  \pp\ data are also shown.
The error bars and shaded bands indicate statistical and systematic
  uncertainties, respectively.
Right panel: the ratio of the \vtwo in 8 and 13~\TeV\ \Zboson-tagged
  samples to the \vtwo in inclusive 13~\TeV\ \pp\ collisions as a
  function of \pt.
The horizontal dotted line indicates unity and is kept to guide the eye.
Results are plotted for the $40<\ntrks\leq100$ multiplicity interval.
\label{fig:template_fits_v2_final_pt}}
\end{figure}

\FloatBarrier
\section{Summary}\label{sec:summary}
In this analysis, the long-range component ($|\deta|>2$) of
  two-particle correlations (2PC) in $\sqs=8$ and 13~\TeV\ \pp\
  collisions containing a \Zboson boson is studied using data collected by the ATLAS detector at the LHC.
The datasets correspond to an integrated luminosity of \luminosity\
  for the 8~\TeV\ data and \luminosityb\ for the 13~\TeV\ data.

The correlations are studied using a template-fitting procedure
  that separates the true long-range correlation from the dijet contribution.
Due to the high-luminosity conditions, a significant contribution to
  the 2PC from pile-up events is observed that contaminates the
  measured correlations.
A new pile-up correction procedure is developed to remove the contribution of pile-up tracks.

The pile-up-corrected 2PC are measured across a large range of
  track multiplicities and over the 0.5--5~\GeV\ \pt range.
The second-order Fourier coefficient of the
  single-particle anisotropy, \vtwo, is extracted and its multiplicity
  and \pt\ dependence is compared to that observed in inclusive \pp\ collisions.
The pile-up-corrected \vtwo values show no significant dependence on the
  event multiplicity, similar to that observed in inclusive \pp\ collisions.
The magnitude of the observed \vtwo as a function of multiplicity and \pt\
  is found to be consistent with that observed in inclusive \pp\ collisions.
These measurements demonstrate that in \pp\ collisions, the long-range
  correlation involving soft particles is not significantly altered
  by the presence of a hard-scattering process.
This result is an important contribution towards   a better understanding of the origin of the long-range
  correlations observed in \pp\ collisions.

\section*{Acknowledgements}

We thank CERN for the very successful operation of the LHC, as well as the
support staff from our institutions without whom ATLAS could not be
operated efficiently.

We acknowledge the support of ANPCyT, Argentina; YerPhI, Armenia; ARC, Australia; BMWFW and FWF, Austria; ANAS, Azerbaijan; SSTC, Belarus; CNPq and FAPESP, Brazil; NSERC, NRC and CFI, Canada; CERN; CONICYT, Chile; CAS, MOST and NSFC, China; COLCIENCIAS, Colombia; MSMT CR, MPO CR and VSC CR, Czech Republic; DNRF and DNSRC, Denmark; IN2P3-CNRS, CEA-DRF/IRFU, France; SRNSFG, Georgia; BMBF, HGF, and MPG, Germany; GSRT, Greece; RGC, Hong Kong SAR, China; ISF and Benoziyo Center, Israel; INFN, Italy; MEXT and JSPS, Japan; CNRST, Morocco; NWO, Netherlands; RCN, Norway; MNiSW and NCN, Poland; FCT, Portugal; MNE/IFA, Romania; MES of Russia and NRC KI, Russian Federation; JINR; MESTD, Serbia; MSSR, Slovakia; ARRS and MIZ\v{S}, Slovenia; DST/NRF, South Africa; MINECO, Spain; SRC and Wallenberg Foundation, Sweden; SERI, SNSF and Cantons of Bern and Geneva, Switzerland; MOST, Taiwan; TAEK, Turkey; STFC, United Kingdom; DOE and NSF, United States of America. In addition, individual groups and members have received support from BCKDF, CANARIE, CRC and Compute Canada, Canada; COST, ERC, ERDF, Horizon 2020, and Marie Sk{\l}odowska-Curie Actions, European Union; Investissements d' Avenir Labex and Idex, ANR, France; DFG and AvH Foundation, Germany; Herakleitos, Thales and Aristeia programmes co-financed by EU-ESF and the Greek NSRF, Greece; BSF-NSF and GIF, Israel; CERCA Programme Generalitat de Catalunya, Spain; The Royal Society and Leverhulme Trust, United Kingdom. 

The crucial computing support from all WLCG partners is acknowledged gratefully, in particular from CERN, the ATLAS Tier-1 facilities at TRIUMF (Canada), NDGF (Denmark, Norway, Sweden), CC-IN2P3 (France), KIT/GridKA (Germany), INFN-CNAF (Italy), NL-T1 (Netherlands), PIC (Spain), ASGC (Taiwan), RAL (UK) and BNL (USA), the Tier-2 facilities worldwide and large non-WLCG resource providers. Major contributors of computing resources are listed in Ref.~\cite{ATL-GEN-PUB-2016-002}.

\printbibliography

@Booklet{ATL-GEN-PUB-2016-002,
          author         = "{ATLAS Collaboration}",
          title          = "{ATLAS Computing Acknowledgements}",
          howpublished   = "{ATL-GEN-PUB-2016-002}",
          url            = "https://cds.cern.ch/record/2202407",
      }

@Article{PERF-2007-01,
    author         = "{ATLAS Collaboration}",
    title          = "{The ATLAS Experiment at the CERN Large Hadron Collider}",
    journal        = "JINST",
    volume         = "3",
    year           = "2008",
    pages          = "S08003",
    doi            = "10.1088/1748-0221/3/08/S08003",
    primaryClass   = "hep-ex",
}

@Article{SOFT-2010-01,
    author         = "{ATLAS Collaboration}",
    title          = "{The ATLAS Simulation Infrastructure}",
    journal        = "Eur. Phys. J. C",
    volume         = "70",
    year           = "2010",
    pages          = "823",
    doi            = "10.1140/epjc/s10052-010-1429-9",
    eprint         = "1005.4568",
    archivePrefix  = "arXiv",
    primaryClass   = "physics.ins-det",
}

@Article{STDM-2010-01,
    author         = "{ATLAS Collaboration}",
    title          = "{Charged-particle multiplicities in \(pp\) interactions at \(\sqrt{s} = 900~\text{GeV}\) measured with the ATLAS detector at the LHC}",
    journal        = "Phys. Lett. B",
    volume         = "688",
    year           = "2010",
    pages          = "21",
    doi            = "10.1016/j.physletb.2010.03.064",
    reportNumber   = "CERN-PH-EP-2010-004",
    eprint         = "1003.3124",
    archivePrefix  = "arXiv",
    primaryClass   = "hep-ex",
}

@Article{HION-2011-01,
    author         = "{ATLAS Collaboration}",
    title          = "{Measurement of the azimuthal anisotropy for charged particle production in \(\sqrt{s_{\text{NN}}}=2.76~\text{TeV}\) lead--lead collisions with the ATLAS detector}",
    journal        = "Phys. Rev. C",
    volume         = "86",
    year           = "2012",
    pages          = "014907",
    doi            = "10.1103/PhysRevC.86.014907",
    reportNumber   = "CERN-PH-EP-2012-035",
    eprint         = "1203.3087",
    archivePrefix  = "arXiv",
    primaryClass   = "hep-ex",
}

@Article{HION-2011-05,
    author         = "{ATLAS Collaboration}",
    title          = "{Measurement of the pseudorapidity and transverse momentum dependence of the elliptic flow of charged particles in lead--lead collisions at \(\sqrt{s_{\text{NN}}}=2.76~\text{TeV}\) with the ATLAS detector}",
    journal        = "Phys. Lett. B",
    volume         = "707",
    year           = "2012",
    pages          = "330",
    doi            = "10.1016/j.physletb.2011.12.056",
    reportNumber   = "CERN-PH-EP-2011-124",
    eprint         = "1108.6018",
    archivePrefix  = "arXiv",
    primaryClass   = "hep-ex",
}

@Article{PERF-2011-02,
    author         = "{ATLAS Collaboration}",
    title          = "{Performance of the ATLAS Trigger System in 2010}",
    journal        = "Eur. Phys. J. C",
    volume         = "72",
    year           = "2012",
    pages          = "1849",
    doi            = "10.1140/epjc/s10052-011-1849-1",
    reportNumber   = "CERN-PH-EP-2011-078",
    eprint         = "1110.1530",
    archivePrefix  = "arXiv",
    primaryClass   = "hep-ex",
}

@Article{STDM-2011-42,
    author         = "{ATLAS Collaboration}",
    title          = "{Measurement of distributions sensitive to the underlying event in inclusive \(Z\)-boson production in \(pp\) collisions at \(\sqrt{s} = 7~\text{TeV}\) with the ATLAS detector}",
    journal        = "Eur. Phys. J. C",
    volume         = "74",
    year           = "2014",
    pages          = "3195",
    doi            = "10.1140/epjc/s10052-014-3195-6",
    reportNumber   = "CERN-PH-EP-2014-162",
    eprint         = "1409.3433",
    archivePrefix  = "arXiv",
    primaryClass   = "hep-ex",
}

@Article{HION-2012-10,
    author         = "{ATLAS Collaboration}",
    title          = "{Measurement of the distributions of event-by-event flow harmonics in lead--lead collisions at \(\sqrt{s_{\text{NN}}} = 2.76~\text{TeV}\) with the ATLAS detector at the LHC}",
    journal        = "JHEP",
    volume         = "11",
    year           = "2013",
    pages          = "183",
    doi            = "10.1007/JHEP11(2013)183",
    reportNumber   = "CERN-PH-EP-2013-04",
    eprint         = "1305.2942",
    archivePrefix  = "arXiv",
    primaryClass   = "hep-ex",
}

@Article{HION-2012-13,
    author         = "{ATLAS Collaboration}",
    title          = "{Observation of Associated Near-Side and Away-Side Long-Range Correlations in \(\sqrt{s_{\text{NN}}} = 5.02~\text{TeV}\) Proton--Lead Collisions with the ATLAS Detector}",
    journal        = "Phys. Rev. Lett.",
    volume         = "110",
    year           = "2013",
    pages          = "182302",
    doi            = "10.1103/PhysRevLett.110.182302",
    reportNumber   = "CERN-PH-EP-2012-366",
    eprint         = "1212.5198",
    archivePrefix  = "arXiv",
    primaryClass   = "hep-ex",
}

@Article{TRIG-2012-03,
    author         = "{ATLAS Collaboration}",
    title          = "{Performance of the ATLAS muon trigger in \(pp\) collisions at \(\sqrt{s} = 8~\text{TeV}\)}",
    journal        = "Eur. Phys. J. C",
    volume         = "75",
    year           = "2015",
    pages          = "120",
    doi            = "10.1140/epjc/s10052-015-3325-9",
    reportNumber   = "CERN-PH-EP-2014-154",
    eprint         = "1408.3179",
    archivePrefix  = "arXiv",
    primaryClass   = "hep-ex",
}

@Article{HION-2013-04,
    author         = "{ATLAS Collaboration}",
    title          = "{Measurement of long-range pseudorapidity correlations and azimuthal harmonics in \(\sqrt{s_{\text{NN}}} = 5.02~\text{TeV}\) proton--lead collisions with the ATLAS detector}",
    journal        = "Phys. Rev. C",
    volume         = "90",
    year           = "2014",
    pages          = "044906",
    doi            = "10.1103/PhysRevC.90.044906",
    reportNumber   = "CERN-PH-EP-2014-201",
    eprint         = "1409.1792",
    archivePrefix  = "arXiv",
    primaryClass   = "hep-ex",
}

@Article{PERF-2014-05,
    author         = "{ATLAS Collaboration}",
    title          = "{Measurement of the muon reconstruction performance of the ATLAS detector using 2011 and 2012 LHC proton--proton collision data}",
    journal        = "Eur. Phys. J. C",
    volume         = "74",
    year           = "2014",
    pages          = "3130",
    doi            = "10.1140/epjc/s10052-014-3130-x",
    reportNumber   = "CERN-PH-EP-2014-151",
    eprint         = "1407.3935",
    archivePrefix  = "arXiv",
    primaryClass   = "hep-ex",
}

@Article{HION-2015-09,
    author         = "{ATLAS Collaboration}",
    title          = "{Observation of Long-Range Elliptic Azimuthal Anisotropies in \(\sqrt{s} = 13\) and \(2.76~\text{TeV}\) \(pp\) Collisions with the ATLAS Detector}",
    journal        = "Phys. Rev. Lett.",
    volume         = "116",
    year           = "2016",
    pages          = "172301",
    doi            = "10.1103/PhysRevLett.116.172301",
    reportNumber   = "CERN-PH-EP-2015-251",
    eprint         = "1509.04776",
    archivePrefix  = "arXiv",
    primaryClass   = "hep-ex",
}

@Article{PERF-2015-10,
    author         = "{ATLAS Collaboration}",
    title          = "{Muon reconstruction performance of the ATLAS detector in proton--proton collision data at \(\sqrt{s} = 13~\text{TeV}\)}",
    journal        = "Eur. Phys. J. C",
    volume         = "76",
    year           = "2016",
    pages          = "292",
    doi            = "10.1140/epjc/s10052-016-4120-y",
    reportNumber   = "CERN-EP-2016-033",
    eprint         = "1603.05598",
    archivePrefix  = "arXiv",
    primaryClass   = "hep-ex",
}

@Article{STDM-2015-02,
    author         = "{ATLAS Collaboration}",
    title          = "{Charged-particle distributions in \(\sqrt{s} = 13~\text{TeV}\) \(pp\) interactions measured with the ATLAS detector at the LHC}",
    journal        = "Phys. Lett. B",
    volume         = "758",
    year           = "2016",
    pages          = "67",
    doi            = "10.1016/j.physletb.2016.04.050",
    reportNumber   = "CERN-EP-2016-014",
    eprint         = "1602.01633",
    archivePrefix  = "arXiv",
    primaryClass   = "hep-ex",
}

@Article{HION-2016-01,
    author         = "{ATLAS Collaboration}",
    title          = "{Measurements of long-range azimuthal anisotropies and associated Fourier coefficients for \(pp\) collisions at \(\sqrt{s}=5.02\) and \(13~\text{TeV}\) and \(p\)+Pb collisions at \(\sqrt{s_{\text{NN}}}=5.02\) TeV with the ATLAS detector}",
    journal        = "Phys. Rev. C",
    volume         = "96",
    year           = "2017",
    pages          = "024908",
    doi            = "10.1103/PhysRevC.96.024908",
    reportNumber   = "CERN-EP-2016-200",
    eprint         = "1609.06213",
    archivePrefix  = "arXiv",
    primaryClass   = "hep-ex",
}

@Article{HION-2016-06,
    author         = "{ATLAS Collaboration}",
    title          = "{Measurement of the azimuthal anisotropy of charged particles produced in \(\sqrt{s_{\text{NN}}} = 5.02~\text{TeV}\) Pb+Pb collisions with the ATLAS detector}",
    journal        = "Eur. Phys. J. C",
    volume         = "78",
    year           = "2018",
    pages          = "997",
    doi            = "10.1140/epjc/s10052-018-6468-7",
    reportNumber   = "CERN-EP-2018-194",
    eprint         = "1808.03951",
    archivePrefix  = "arXiv",
    primaryClass   = "hep-ex",
}

@Article{TRIG-2016-01,
    author         = "{ATLAS Collaboration}",
    title          = "{Performance of the ATLAS trigger system in 2015}",
    journal        = "Eur. Phys. J. C",
    volume         = "77",
    year           = "2017",
    pages          = "317",
    doi            = "10.1140/epjc/s10052-017-4852-3",
    reportNumber   = "CERN-EP-2016-241",
    eprint         = "1611.09661",
    archivePrefix  = "arXiv",
    primaryClass   = "hep-ex",
}

@Article{HION-2019-03, 
    author         = "{ATLAS Collaboration}", 
    title          = "{Measurement of the azimuthal anisotropy of charged-particle production in Xe+Xe collisions at $\sqrt{s_{\mathrm{NN}}}=5.44$ TeV with the ATLAS detector}",
    year           = "2019",
    reportNumber   = "CERN-EP-2018-100", 
    eprint         = "1911.04812",
    archivePrefix  = "arXiv",
    primaryClass   = "nucl-ex"
}

@Article{CMS-HIN-10-002,
    author         = "{CMS Collaboration}",
    title          = "{Measurement of the elliptic anisotropy of charged particles produced in PbPb collisions at $\sqn=2.76$ TeV}",
    journal        = "Phys. Rev. C",
    volume         = "87",
    year           = "2013",
    pages          = "014902",
    doi            = "10.1103/PhysRevC.87.014902",
    reportNumber   = "CERN-PH-EP-2012-095",
    eprint         = "1204.1409",
    archivePrefix  = "arXiv",
    primaryClass   = "hep-ex",
}

@Article{CMS-QCD-10-002,
    author         = "{CMS Collaboration}",
    title          = "{Observation of long-range, near-side angular correlations in proton--proton collisions at the LHC}",
    journal        = "JHEP",
    volume         = "09",
    year           = "2010",
    pages          = "091",
    doi            = "10.1007/JHEP09(2010)091",
    reportNumber   = "CERN-PH-EP-2010-031",
    eprint         = "1009.4122",
    archivePrefix  = "arXiv",
    primaryClass   = "hep-ex",
}

@Article{CMS-HIN-12-005,
    author         = "{CMS Collaboration}",
    title          = "{Observation of long-range, near-side angular correlations in $p$Pb collisions at the LHC}",
    journal        = "Phys. Lett. B",
    volume         = "718",
    year           = "2013",
    pages          = "795",
    doi            = "10.1016/j.physletb.2012.11.025",
    reportNumber   = "CERN-PH-EP-2012-320",
    eprint         = "1210.5482",
    archivePrefix  = "arXiv",
    primaryClass   = "hep-ex",
}

@Article{CMS-HIN-16-010,
    author         = "{CMS Collaboration}",
    title          = "{Evidence for collectivity in $pp$ collisions at the LHC}",
    journal        = "Phys. Lett. B",
    volume         = "765",
    year           = "2017",
    pages          = "193",
    doi            = "10.1016/j.physletb.2016.12.009",
    reportNumber   = "CERN-EP-2016-147",
    eprint         = "1606.06198",
    archivePrefix  = "arXiv",
    primaryClass   = "hep-ex",
}

@Booklet{ATL-PHYS-PUB-2012-003,
    author         = "{ATLAS Collaboration}",
    title          = "{Summary of ATLAS Pythia 8 tunes}",
    howpublished   = "{ATL-PHYS-PUB-2012-003}",
    url            = "https://cds.cern.ch/record/1474107",
    year           = "2012",
}

@Booklet{ATL-PHYS-PUB-2015-018,
    author         = "{ATLAS Collaboration}",
    title          = "{Track Reconstruction Performance of the ATLAS Inner Detector at  $\sqrt{s} = 13\;\mbox{TeV}$}",
    howpublished   = "{ATL-PHYS-PUB-2015-018}",
    url            = "https://cds.cern.ch/record/2037683",
    year           = "2015",
}

@article{Adams:2005dq, 
  Archiveprefix = {arXiv}, 
  Author = {{STAR Collaboration,  J.~Adams et al.}}, 
  Collaboration = {STAR}, 
  Doi = {10.1016/j.nuclphysa.2005.03.085}, 
  Eprint = {nucl-ex/0501009}, 
  Journal = {Nucl. Phys. A}, 
  Pages = {102-183}, 
  Primaryclass = {nucl-ex}, 
  Slaccitation = {%%CITATION = NUCL-EX/0501009;%%}, 
  Title = {{Experimental and theoretical challenges 
    in the search for the quark gluon plasma: 
    The STAR Collaboration's critical assessment of the evidence from RHIC collisions}}, 
  Volume = {757}, 
  Year = {2005}
}

@article{Back:2004je, 
  Archiveprefix = {arXiv}, 
  Author = {{PHOBOS Collaboration,  B. B. Back et al.}}, 
  Collaboration = {PHOBOS}, 
  Doi = {10.1016/j.nuclphysa.2005.03.084}, 
  Eprint = {nucl-ex/0410022}, 
  Journal = {Nucl. Phys. A}, 
  Pages = {28-101}, 
  Primaryclass = {nucl-ex}, 
  Slaccitation = {%%CITATION = NUCL-EX/0410022;%%}, 
  Title = {{The PHOBOS perspective on discoveries at RHIC}}, 
  Volume = {757}, 
  Year = {2005}
}

@article{Adcox:2004mh, 
  Archiveprefix = {arXiv}, 
  Author = {{PHENIX Collaboration,  K. Adcox et al.}}, 
  Collaboration = {PHENIX},
  Doi = {10.1016/j.nuclphysa.2005.03.086},
  Eprint = {nucl-ex/0410003},
  Journal = {Nucl. Phys. A},
  Pages = {184-283},
  Primaryclass = {nucl-ex},
  Slaccitation = {%%CITATION = NUCL-EX/0410003;%%},
  Title = {{Formation of dense partonic matter in relativistic 
    nucleus-nucleus collisions at RHIC: Experimental evaluation by the PHENIX Collaboration}}, 
  Volume = {757},
  Year = {2005}
}

@article{Arsene:2004fa, 
  Archiveprefix = {arXiv}, 
  Author = {{BRAHMS Collaboration,  I.~Arsene et al.}}, 
  Collaboration = {BRAHMS}, 
  Doi = {10.1016/j.nuclphysa.2005.02.130}, 
  Eprint = {nucl-ex/0410020}, 
  Journal = {Nucl. Phys. A}, 
  Pages = {1-27}, 
  Primaryclass = {nucl-ex}, 
  Slaccitation = {%%CITATION = NUCL-EX/0410020;%%}, 
  Title = {{Quark--gluon plasma and color glass condensate at RHIC? 
            The Perspective from the BRAHMS experiment}}, 
  Volume = {757}, 
  Year = {2005}
}

@article{He:2015hfa, 
      author         = "He,  Liang and Edmonds,  Terrence and Lin,  Zi-Wei and Liu, 
                        Feng and Molnar,  Denes and Wang,  Fuqiang", 
      title          = "{Anisotropic parton escape is the dominant source of
                        azimuthal anisotropy in transport models}", 
      journal        = "Phys. Lett.", 
      volume         = "B753", 
      year           = "2016", 
      pages          = "506-510", 
      doi            = "10.1016/j.physletb.2015.12.051", 
      eprint         = "1502.05572", 
      archivePrefix  = "arXiv", 
      primaryClass   = "nucl-th", 
      SLACcitation   = "%%CITATION = ARXIV:1502.05572;%%"
}

@article{Romatschke:2015dha, 
      author         = "Romatschke,  Paul", 
      title          = "{Collective flow without hydrodynamics: simulation
                        results for relativistic ion collisions}", 
      journal        = "Eur. Phys. J.", 
      volume         = "C75", 
      year           = "2015", 
      number         = "9", 
      pages          = "429", 
      doi            = "10.1140/epjc/s10052-015-3646-8", 
      eprint         = "1504.02529", 
      archivePrefix  = "arXiv", 
      primaryClass   = "nucl-th", 
      SLACcitation   = "%%CITATION = ARXIV:1504.02529;%%"
}

@article{Kurkela:2018ygx, 
      author         = "Kurkela,  Aleksi and Wiedemann,  Urs Achim and Wu,  Bin", 
      title          = "{Nearly isentropic flow at sizeable $\eta/s$}", 
      journal        = "Phys. Lett.", 
      volume         = "B783", 
      year           = "2018", 
      pages          = "274-279", 
      doi            = "10.1016/j.physletb.2018.06.064", 
      eprint         = "1803.02072", 
      archivePrefix  = "arXiv", 
      primaryClass   = "hep-ph", 
      reportNumber   = "CERN-TH-2018-045", 
      SLACcitation   = "%%CITATION = ARXIV:1803.02072;%%"
}

@article{Gleisberg:2008ta,
      author         = "Gleisberg,  T. and Hoeche,  Stefan. and Krauss,  F. and
                        Schonherr,  M. and Schumann,  S. and Siegert,  F. and Winter,
                        J.",
      title          = "{Event generation with SHERPA 1.1}",
      journal        = "JHEP",
      volume         = "02",
      year           = "2009",
      pages          = "007",
      doi            = "10.1088/1126-6708/2009/02/007",
      eprint         = "0811.4622",
      archivePrefix  = "arXiv",
      primaryClass   = "hep-ph",
      reportNumber   = "FERMILAB-PUB-08-477-T,  SLAC-PUB-13420,  ZU-TH-17-08,
                        DCPT-08-138,  IPPP-08-69,  EDINBURGH-2008-30,  MCNET-08-14",
      SLACcitation   = "%%CITATION = ARXIV:0811.4622;%%"
}

@article{Snellings:2011sz,
      author         = "Snellings,  Raimond",
      title          = "{Elliptic Flow: A Brief Review}",
      journal        = "New J. Phys.",
      volume         = "13",
      year           = "2011",
      pages          = "055008",
      doi            = "10.1088/1367-2630/13/5/055008",
      eprint         = "1102.3010",
      archivePrefix  = "arXiv",
      primaryClass   = "nucl-ex",
      SLACcitation   = "%%CITATION = ARXIV:1102.3010;%%"
}

@article{Gale:2013da,
      author         = "Gale,  Charles and Jeon,  Sangyong and Schenke,  Bjoern",
      title          = "{Hydrodynamic Modeling of Heavy-Ion Collisions}",
      journal        = "Int. J. Mod. Phys. A",
      volume         = "28",
      year           = "2013",
      pages          = "1340011",
      doi            = "10.1142/S0217751X13400113",
      eprint         = "1301.5893",
      archivePrefix  = "arXiv",
      primaryClass   = "nucl-th",
      SLACcitation   = "%%CITATION = ARXIV:1301.5893;%%"
}

@article{Heinz:2013th, 
      author         = "Heinz,  Ulrich and Snellings,  Raimond", 
      title          = "{Collective Flow and Viscosity in Relativistic Heavy-Ion Collisions}", 
      journal        = "Ann.~Rev.~Nucl.~Part.~Sci.", 
      volume         = "63", 
      pages          = "123-151", 
      doi            = "10.1146/annurev-nucl-102212-170540", 
      year           = "2013", 
      eprint         = "1301.2826", 
      archivePrefix  = "arXiv", 
      primaryClass   = "nucl-th", 
      SLACcitation   = "%%CITATION = ARXIV:1301.2826;%%", 
}

@article{Frankfurt:2003td,
      author         = "Frankfurt,  L. and Strikman,  M. and Weiss,  C.",
      title          = "{Dijet production as a centrality trigger for $pp$
                        collisions at CERN LHC}",
      journal        = "Phys. Rev. D",
      volume         = "69",
      year           = "2004",
      pages          = "114010",
      doi            = "10.1103/PhysRevD.69.114010",
      eprint         = "hep-ph/0311231",
      archivePrefix  = "arXiv",
      primaryClass   = "hep-ph",
      SLACcitation   = "%%CITATION = HEP-PH/0311231;%%"
}

@article{Abelev:2012ola,
      author         = "{ALICE Collaboration}",
      title          = "{Long-range angular correlations on the near and away
                        side in \pPb collisions at $\sqn=5.02$ TeV}",
      journal        = "Phys.~Lett. B",
      volume         = "719",
      year           = "2013",
      pages          = "29-41",
      doi            = "10.1016/j.physletb.2013.01.012",
      eprint         = "1212.2001",
      archivePrefix  = "arXiv",
      primaryClass   = "nucl-ex",
      reportNumber   = "CERN-PH-EP-2012-359",
      SLACcitation   = "%%CITATION = ARXIV:1212.2001;%%"
}

@article{Aamodt:2010pa,
        Archiveprefix = {arXiv},
        Author = {{{ALICE Collaboration}}},
        Collaboration = {ALICE},
        Doi = {10.1103/PhysRevLett.105.252302},
        Eprint = {1011.3914},
        Journal = {Phys. Rev. Lett.},
        Pages = {252302},
        Primaryclass = {nucl-ex},
        Slaccitation = {%%CITATION = ARXIV:1011.3914;%%},
        Title = {{Elliptic Flow of Charged Particles in Pb-Pb Collisions at $\sqn = 2.76$ TeV}},
        Volume = {105},
        Year = {2010}
}

@article{Sorge:1996pc,
      author         = "Sorge, H.",
      title          = "{Elliptical Flow: A Signature for Early Pressure in
                        Ultrarelativistic Nucleus-Nucleus Collisions}",
      journal        = "Phys. Rev. Lett.",
      volume         = "78",
      year           = "1997",
      pages          = "2309-2312",
      doi            = "10.1103/PhysRevLett.78.2309",
      eprint         = "nucl-th/9610026",
      archivePrefix  = "arXiv",
      xxprimaryClass   = "nucl-th",
      reportNumber   = "SUNY-NTG-96-40",
      SLACcitation   = "%%CITATION = NUCL-TH/9610026;%%"
}

@article{Kolb:2001qz,
      author         = "Kolb, P. F. and Heinz, Ulrich W. and Huovinen, P. and
                        Eskola, K. J. and Tuominen, Kimmo",
      title          = "{Centrality dependence of multiplicity, transverse
                        energy, and elliptic flow from hydrodynamics}",
      journal        = "Nucl. Phys. A",
      volume         = "696",
      year           = "2001",
      pages          = "197-215",
      doi            = "10.1016/S0375-9474(01)01114-9",
      eprint         = "hep-ph/0103234",
      archivePrefix  = "arXiv",
      xxprimaryClass   = "hep-ph",
      reportNumber   = "JYFL-4-2001, LBNL-47635",
      SLACcitation   = "%%CITATION = HEP-PH/0103234;%%"
}

@article{Huovinen:2001cy,
      author         = "Huovinen, P. and Kolb, P. F. and Heinz, Ulrich W. and
                        Ruuskanen, P. V. and Voloshin, S. A.",
      title          = "{Radial and elliptic flow at RHIC: Further predictions}",
      journal        = "Phys. Lett. B",
      volume         = "503",
      year           = "2001",
      pages          = "58-64",
      doi            = "10.1016/S0370-2693(01)00219-2",
      eprint         = "hep-ph/0101136",
      archivePrefix  = "arXiv",
      xxprimaryClass   = "hep-ph",
      SLACcitation   = "%%CITATION = HEP-PH/0101136;%%"
}

@article{Welsh:2016siu,
      author         = "Welsh, Kevin and Singer, Jordan and Heinz, Ulrich W.",
      title          = "{Initial state fluctuations in collisions between light
                        and heavy ions}",
      journal        = "Phys. Rev. C",
      volume         = "94",
      year           = "2016",
      number         = "2",
      pages          = "024919",
      doi            = "10.1103/PhysRevC.94.024919",
      eprint         = "1605.09418",
      archivePrefix  = "arXiv",
      xprimaryClass   = "nucl-th",
      SLACcitation   = "%%CITATION = ARXIV:1605.09418;%%"
}

@article{Agostinelli:2002hh,
      author         = "Agostinelli, S. and others",
      title          = "{Geant4: a simulation toolkit}",
      collaboration  = "GEANT4",
      journal        = "Nucl. Instrum. Meth. A",
      volume         = "506",
      year           = "2003",
      pages          = "250-303",
      doi            = "10.1016/S0168-9002(03)01368-8",
      reportNumber   = "SLAC-PUB-9350, FERMILAB-PUB-03-339",
      SLACcitation   = "%%CITATION = NUIMA,A506,250;%%"
}

@article{Dumitru:2010iy,
      author         = "{A. Dumitru et al.}",
      title          = "{The Ridge in proton-proton collisions at the LHC}",
      journal        = "Phys.Lett. B",
      volume         = "697",
      pages          = "21-25",
      doi            = "10.1016/j.physletb.2011.01.024",
      year           = "2011",
      eprint         = "1009.5295",
      archivePrefix  = "arXiv",
      xprimaryClass   = "hep-ph",
      reportNumber   = "INT-PUB-10-051, BCCUNY-HEP-10-03, BNL-94103-2010-JA,
                        RBRC-858",
      SLACcitation   = "%%CITATION = ARXIV:1009.5295;%%",
}

@article{Kovner:2010xk,
      author         = "Kovner, Alex and Lublinsky, Michael",
      title          = "{Angular Correlations in Gluon Production at High
                        Energy}",
      journal        = "Phys. Rev. D",
      volume         = "83",
      pages          = "034017",
      doi            = "10.1103/PhysRevD.83.034017",
      year           = "2011",
      eprint         = "1012.3398",
      archivePrefix  = "arXiv",
      xprimaryClass   = "hep-ph",
      SLACcitation   = "%%CITATION = ARXIV:1012.3398;%%",
}

@article{Altinoluk:2011qy,
      author         = "Altinoluk, T. and Kovner, A.",
      title          = "{Particle production at high energy and large transverse
                        momentum - 'The hybrid formalism' revisited}",
      journal        = "Phys. Rev. D",
      volume         = "83",
      pages          = "105004",
      doi            = "10.1103/PhysRevD.83.105004",
      year           = "2011",
      eprint         = "1102.5327",
      archivePrefix  = "arXiv",
      xprimaryClass   = "hep-ph",
      SLACcitation   = "%%CITATION = ARXIV:1102.5327;%%",
}

@article{Dusling:2012iga,
      author         = "Dusling, Kevin and Venugopalan, Raju",
      title          = "{Azimuthal Collimation of Long Range Rapidity
                        Correlations by Strong Color Fields in High Multiplicity
                        Hadron-Hadron Collisions}",
      journal        = "Phys. Rev. Lett.",
      volume         = "108",
      pages          = "262001",
      doi            = "10.1103/PhysRevLett.108.262001",
      year           = "2012",
      eprint         = "1201.2658",
      archivePrefix  = "arXiv",
      xprimaryClass   = "hep-ph",
      SLACcitation   = "%%CITATION = ARXIV:1201.2658;%%",
}

@article{Levin:2011fb,
      author         = "Levin, Eugene and Rezaeian, Amir H.",
      title          = "{The Ridge from the BFKL evolution and beyond}",
      journal        = "Phys. Rev. D",
      volume         = "84",
      pages          = "034031",
      doi            = "10.1103/PhysRevD.84.034031",
      year           = "2011",
      eprint         = "1105.3275",
      archivePrefix  = "arXiv",
      xprimaryClass   = "hep-ph",
      SLACcitation   = "%%CITATION = ARXIV:1105.3275;%%",
}

@article{Strikman:2011cx,
      author         = "Strikman, Mark",
      title          = "{Transverse Nucleon Structure and Multiparton
                        Interactions}",
      journal        = "Acta Phys. Polon. B",
      volume         = "42",
      pages          = "2607-2630",
      doi            = "10.5506/APhysPolB.42.2607",
      year           = "2011",
      eprint         = "1112.3834",
      archivePrefix  = "arXiv",
      xprimaryClass   = "hep-ph",
      SLACcitation   = "%%CITATION = ARXIV:1112.3834;%%",
}

@article{Qin:2010pf,
      author         = "Qin, Guang-You and Petersen, Hannah and Bass, Steffen A.
                        and Muller, Berndt",
      title          = "{Translation of collision geometry fluctuations into
                        momentum anisotropies in relativistic heavy-ion
                        collisions}",
      journal        = "Phys. Rev. C",
      volume         = "82",
      year           = "2010",
      pages          = "064903",
      doi            = "10.1103/PhysRevC.82.064903",
      eprint         = "1009.1847",
      archivePrefix  = "arXiv",
      primaryClass   = "nucl-th",
      SLACcitation   = "%%CITATION = ARXIV:1009.1847;%%"
}

@article{Aaboud:2016rmg,
      author         = "{ATLAS Collaboration}",
      xauthor         = "Aaboud, Morad and others",
      title          = "{Reconstruction of primary vertices at the ATLAS
                        experiment in Run 1 proton-proton collisions at the
                        LHC}",
      collaboration  = "ATLAS",
      journal        = "Eur. Phys. J. C",
      volume         = "77",
      year           = "2017",
      number         = "5",
      pages          = "332",
      doi            = "10.1140/epjc/s10052-017-4887-5",
      eprint         = "1611.10235",
      archivePrefix  = "arXiv",
      primaryClass   = "physics.ins-det",
      reportNumber   = "CERN-EP-2016-150",
      SLACcitation   = "%%CITATION = ARXIV:1611.10235;%%"
}

@article{Abbott:2018ikt,
      author         = "Abbott,  B. and others",
      title          = "{Production and Integration of the ATLAS Insertable
                        B-Layer}",
      collaboration  = "ATLAS IBL",
      journal        = "JINST",
      volume         = "13",
      year           = "2018",
      number         = "05",
      pages          = "T05008",
      doi            = "10.1088/1748-0221/13/05/T05008",
      eprint         = "1803.00844",
      archivePrefix  = "arXiv",
      primaryClass   = "physics.ins-det",
      SLACcitation   = "%%CITATION = ARXIV:1803.00844;%%"
}

@Booklet{ATLAS-TDR-2010-19,
    author         = "{ATLAS Collaboration}",
    title          = "{ATLAS Insertable B-Layer Technical Design Report}",
    howpublished   = "ATLAS-TDR-19",
    url            = "https://cds.cern.ch/record/1291633",
    year           = "2010",
    related        = "ATLAS-TDR-2010-19-add",
    relatedstring  = "Addendum:",
}

@article{Aaboud:2016hhf,
      author         = "{ATLAS Collaboration}",
      xauthor         = "Aaboud, Morad and others",
      title          = "{Luminosity determination in pp collisions at $\sqrt{s}$
                        = 8 TeV using the ATLAS detector at the LHC}",
      collaboration  = "ATLAS",
      journal        = "Eur. Phys. J. C",
      volume         = "76",
      year           = "2016",
      number         = "12",
      pages          = "653",
      doi            = "10.1140/epjc/s10052-016-4466-1",
      eprint         = "1608.03953",
      archivePrefix  = "arXiv",
      primaryClass   = "hep-ex",
      reportNumber   = "CERN-EP-2016-117",
      SLACcitation   = "%%CITATION = ARXIV:1608.03953;%%"
}

@techreport{ATLAS-CONF-2019-021,
      author         = "{ATLAS Collaboration}",
      title         = "{Luminosity determination in $pp$ collisions at
                       $\sqrt{s}=13$ TeV using the ATLAS detector at the LHC}",
      institution   = "CERN",
      collaboration = "ATLAS Collaboration",
      address       = "Geneva",
      number        = "ATLAS-CONF-2019-021",
      month         = "Jun",
      year          = "2019",
      reportNumber  = "ATLAS-CONF-2019-021",
      url           = "http://cds.cern.ch/record/2677054",
}

@article{Sjostrand:2007gs,
      author         = "Sjostrand,  Torbjorn and Mrenna,  Stephen and Skands,  Peter
                        Z.",
      title          = "{A Brief Introduction to PYTHIA 8.1}",
      journal        = "Comput. Phys. Commun.",
      volume         = "178",
      year           = "2008",
      pages          = "852-867",
      doi            = "10.1016/j.cpc.2008.01.036",
      eprint         = "0710.3820",
      archivePrefix  = "arXiv",
      primaryClass   = "hep-ph",
      reportNumber   = "CERN-LCGAPP-2007-04,  LU-TP-07-28,
                        FERMILAB-PUB-07-512-CD-T",
      SLACcitation   = "%%CITATION = ARXIV:0710.3820;%%"
}

@article{Martin:2009iq,
      author         = "Martin,  A. D. and Stirling,  W. J. and Thorne,  R. S. and
                        Watt,  G.",
      title          = "{Parton distributions for the LHC}",
      journal        = "Eur. Phys. J.",
      volume         = "C63",
      year           = "2009",
      pages          = "189-285",
      doi            = "10.1140/epjc/s10052-009-1072-5",
      eprint         = "0901.0002",
      archivePrefix  = "arXiv",
      primaryClass   = "hep-ph",
      reportNumber   = "IPPP-08-95,  DCPT-08-190,  CAVENDISH-HEP-08-16",
      SLACcitation   = "%%CITATION = ARXIV:0901.0002;%%"
}

@article{Acharya:2019vdf,
      author         = "{ALICE Collaboration}",
      title          = "{Investigations of anisotropic flow using multi-particle
                        azimuthal correlations in pp, p-Pb, Xe-Xe, and Pb-Pb
                        collisions at the LHC}",
      journal        = "Phys. Rev. Lett.",
      volume         = "123",
      year           = "2019",
      number         = "14",
      pages          = "142301",
      doi            = "10.1103/PhysRevLett.123.142301",
      eprint         = "1903.01790",
      archivePrefix  = "arXiv",
      primaryClass   = "nucl-ex",
      reportNumber   = "CERN-EP-2019-033",
      SLACcitation   = "%%CITATION = ARXIV:1903.01790;%%"
}

@article{Aaij:2015qcq,
      author         = "{LHCb Collaboration}",
      title          = "{Measurements of long-range near-side angular
                        correlations in $\sqrt{s_{\text{NN}}}=5$TeV proton-lead
                        collisions in the forward region}",
      collaboration  = "LHCb",
      journal        = "Phys. Lett. B",
      volume         = "762",
      year           = "2016",
      pages          = "473-483",
      doi            = "10.1016/j.physletb.2016.09.064",
      eprint         = "1512.00439",
      archivePrefix  = "arXiv",
      primaryClass   = "nucl-ex",
      reportNumber   = "LHCB-PAPER-2015-040, CERN-PH-EP-2015-308",
      SLACcitation   = "%%CITATION = ARXIV:1512.00439;%%"
}
\clearpage 
 
\begin{flushleft}
{\Large The ATLAS Collaboration}

\bigskip

M.~Aaboud$^\textrm{\scriptsize 35d}$,    
G.~Aad$^\textrm{\scriptsize 100}$,    
B.~Abbott$^\textrm{\scriptsize 127}$,    
D.C.~Abbott$^\textrm{\scriptsize 101}$,    
O.~Abdinov$^\textrm{\scriptsize 13,*}$,    
B.~Abeloos$^\textrm{\scriptsize 131}$,    
D.K.~Abhayasinghe$^\textrm{\scriptsize 92}$,    
S.H.~Abidi$^\textrm{\scriptsize 166}$,    
O.S.~AbouZeid$^\textrm{\scriptsize 40}$,    
N.L.~Abraham$^\textrm{\scriptsize 155}$
% ...
\bigskip
\\

% ...
$^{13}$Institute of Physics, Azerbaijan Academy of Sciences, Baku; Azerbaijan.\\
% ...
$^{35}$$^{(a)}$Facult\'e des Sciences Ain Chock, R\'eseau Universitaire de Physique des Hautes Energies - Universit\'e Hassan II, Casablanca;$^{(b)}$Facult\'{e} des Sciences, Universit\'{e} Ibn-Tofail, K\'{e}nitra;$^{(c)}$Facult\'e des Sciences Semlalia, Universit\'e Cadi Ayyad, LPHEA-Marrakech;$^{(d)}$Facult\'e des Sciences, Universit\'e Mohamed Premier and LPTPM, Oujda;$^{(e)}$Facult\'e des sciences, Universit\'e Mohammed V, Rabat; Morocco.\\
% ...
$^{40}$Niels Bohr Institute, University of Copenhagen, Copenhagen; Denmark.\\
% ...
$^{92}$Department of Physics, Royal Holloway University of London, Egham; United Kingdom.\\
% ...
$^{100}$CPPM, Aix-Marseille Universit\'e, CNRS/IN2P3, Marseille; France.\\
$^{101}$Department of Physics, University of Massachusetts, Amherst MA; United States of America.\\
% ...
$^{127}$Homer L. Dodge Department of Physics and Astronomy, University of Oklahoma, Norman OK; United States of America.\\
% ...
$^{131}$LAL, Universit\'e Paris-Sud, CNRS/IN2P3, Universit\'e Paris-Saclay, Orsay; France.\\
% ...
$^{155}$Department of Physics and Astronomy, University of Sussex, Brighton; United Kingdom.\\
% ...
$^{166}$Department of Physics, University of Toronto, Toronto ON; Canada.\\
% ...
$^{a}$ Also at Borough of Manhattan Community College, City University of New York, New York NY; United States of America.\\
$^{b}$ Also at Centre for High Performance Computing, CSIR Campus, Rosebank, Cape Town; South Africa.\\
$^{c}$ Also at CERN, Geneva; Switzerland.\\
$^{d}$ Also at CPPM, Aix-Marseille Universit\'e, CNRS/IN2P3, Marseille; France.\\
$^{e}$ Also at D\'epartement de Physique Nucl\'eaire et Corpusculaire, Universit\'e de Gen\`eve, Gen\`eve; Switzerland.\\
% ...
$^{*}$ Deceased

\end{flushleft}



\end{document}